\documentclass[twocolumn]{aastex631}
\usepackage{CJK}
\usepackage{hyperref}
\usepackage{comment}

\usepackage{lineno, graphicx}
\newcommand{\Lya}{\hbox{{\rm Ly}$\,\alpha$}}

\newcommand{\Hb}{\hbox{{\rm H}$\beta$}}

\newcommand{\HeII}{\hbox{{\rm He}\kern 0.1em{\sc ii}}}
\newcommand{\OII}{\hbox{{\rm [O}\kern 0.1em{\sc ii}{\rm ]}}}
\newcommand{\OIII}{\hbox{{\rm [O}\kern 0.1em{\sc iii}{\rm ]}}}
\newcommand{\CIV}{\hbox{{\rm C}\kern 0.1em{\sc iv}}}
\newcommand{\OVI}{\hbox{{\rm O}\kern 0.1em{\sc vi}}}
\newcommand{\NV}{\hbox{{\rm N}\kern 0.1em{\sc v}}}
\newcommand{\CIII}{\hbox{{\rm C}\kern 0.1em{\sc iii}{\rm ]}}}
\newcommand{\MgII}{\hbox{{\rm Mg}\kern 0.1em{\sc ii}}}
\newcommand{\SiIV}{\hbox{{\rm Si}\kern 0.1em{\sc iv}}}
\newcommand{\FeII}{\hbox{{\rm Fe}\kern 0.1em{\sc ii}}}
\newcommand{\Mbh}{\ensuremath{M_{\rm BH}}}

\newcommand{\pyccf}{\texttt{PyCCF}}
\newcommand{\jav}{\texttt{JAVELIN}}
\newcommand{\hst}{HST}
\newcommand{\mrk}{Mrk\,817}

\begin{document}
\begin{CJK*}{UTF8}{gbsn}

\title{AGN STORM 2: II. Ultraviolet Observations of \mrk\ with the Cosmic Origins Spectrograph on the Hubble Space Telescope\footnote{Based on observations made with the NASA/ESA Hubble Space Telescope, obtained at the Space Telescope Science Institute, which is operated by the Association of Universities for Research in Astronomy, Inc., under NASA contract NAS5-26555. These observations are associated with program GO-16196.”}}

%
%


\author[0000-0002-0957-7151]{Y. Homayouni}
\affiliation{Space Telescope Science Institute, 3700 San Martin Drive, Baltimore, MD 21218, USA}
\affiliation{Department of Astronomy and Astrophysics, The Pennsylvania State University, 525 Davey Laboratory, University Park, PA 16802}
\affiliation{Institute for Gravitation and the Cosmos, The Pennsylvania State University, University Park, PA 16802}

\author[0000-0003-3242-7052]{Gisella De~Rosa}
\affiliation{Space Telescope Science Institute, 3700 San Martin Drive, Baltimore, MD 21218, USA}

\author[0000-0002-2509-3878]{Rachel Plesha}
\affiliation{Space Telescope Science Institute, 3700 San Martin Drive, Baltimore, MD 21218, USA}

\author[0000-0002-2180-8266]{Gerard A.\ Kriss}
\affiliation{Space Telescope Science Institute, 3700 San Martin Drive, Baltimore, MD 21218, USA}

\author[0000-0002-3026-0562]{Aaron J.\ Barth}
\affiliation{Department of Physics and Astronomy, 4129 Frederick Reines Hall, University of California, Irvine, CA, 92697-4575, USA}

\author[0000-0002-8294-9281]{Edward M.\ Cackett}
\affiliation{Department of Physics and Astronomy, Wayne State University, 666 W.\ Hancock St, Detroit, MI, 48201, USA}

\author[0000-0003-1728-0304]{Keith Horne}
\affiliation{SUPA School of Physics and Astronomy, North Haugh, St.~Andrews, KY16~9SS, Scotland, UK}

\author[0000-0003-0172-0854]{Erin A.\ Kara}
\affiliation{MIT Kavli Institute for Astrophysics and Space Research, Massachusetts Institute of Technology, Cambridge, MA 02139, USA}

\author{Hermine Landt}
\affiliation{Centre for Extragalactic Astronomy, Department of Physics, Durham University, South Road, Durham DH1 3LE, UK}


\author[0000-0003-2991-4618]{Nahum Arav}
\affiliation{Department of Physics, Virginia Tech, Blacksburg, VA 24061, USA}


\author[0000-0001-6301-570X]{Benjamin D. Boizelle}
\affiliation{Department of Physics and Astronomy, N284 ESC, Brigham Young University, Provo, UT, 84602, USA}

\author[0000-0002-2816-5398]{Misty C.\ Bentz}
\affiliation{Department of Physics and Astronomy, Georgia State University, 25 Park Place, Suite 605, Atlanta, GA 30303, USA}

\author[0000-0001-5955-2502]{Thomas G. Brink}
\affiliation{Department of Astronomy, University of California, Berkeley, CA 94720-3411, USA}

\author[0000-0002-1207-0909]{Michael S.\ Brotherton}
\affiliation{Department of Physics and Astronomy, University of Wyoming, Laramie, WY 82071, USA}


\author{Doron Chelouche}
\affiliation{Department of Physics, Faculty of Natural Sciences, University of Haifa, Haifa 3498838, Israel}
\affiliation{Haifa Research Center for Theoretical Physics and Astrophysics, University of Haifa, Haifa 3498838, Israel}



\author[0000-0001-9931-8681]{Elena Dalla Bont\`{a}}
\affiliation{Dipartimento di Fisica e Astronomia ``G.\  Galilei,'' Universit\`{a} di Padova, Vicolo dell'Osservatorio 3, I-35122 Padova, Italy}
\affiliation{INAF-Osservatorio Astronomico di Padova, Vicolo dell'Osservatorio 5 I-35122, Padova, Italy}

\author[0000-0002-0964-7500]{Maryam Dehghanian}
\affiliation{Department of Physics and Astronomy, The University of Kentucky, Lexington, KY 40506, USA}
\affiliation{Department of Physics, Virginia Tech, Blacksburg, VA 24061, USA}

\author[0000-0002-5830-3544]{Pu Du} 
\affiliation{Key Laboratory for Particle Astrophysics, Institute of High Energy Physics, Chinese Academy of Sciences, 19B Yuquan Road,\\ Beijing 100049, People's Republic of China}
\author[0000-0003-4503-6333]{Gary J.\ Ferland}
\affiliation{Department of Physics and Astronomy, The University of Kentucky, Lexington, KY 40506, USA}

\author[0000-0002-8224-1128]{Laura Ferrarese}
\affiliation{NRC Herzberg Astronomy and Astrophysics Research Centre, 5071 West Saanich Road, Victoria, BC, V9E 2E7, Canada}

\author[0000-0002-2306-9372]{Carina Fian}
\affiliation{Haifa Research Center for Theoretical Physics and Astrophysics, University of Haifa, Haifa 3498838, Israel}
\affiliation{School of Physics and Astronomy and Wise observatory, Tel Aviv University, Tel Aviv 6997801, Israel}

\author[0000-0003-3460-0103]{Alexei V.\ Filippenko}
\affiliation{Department of Astronomy, University of California, Berkeley, CA 94720-3411, USA}
\affiliation{Miller Institute for Basic Research in Science, University of California, Berkeley, CA 94720, USA}

\author[0000-0002-3365-8875]{Travis Fischer}
\affiliation{Space Telescope Science Institute, 3700 San Martin Drive, Baltimore, MD 21218, USA}

\author[0000-0002-2445-5275]{Ryan J.\ Foley}
\affiliation{Department of Astronomy and Astrophysics, University of California, Santa Cruz, CA 92064, USA}

\author[0000-0001-9092-8619]{Jonathan Gelbord}
\affiliation{Spectral Sciences Inc., 4 Fourth Ave., Burlington, MA 01803, USA}

\author[0000-0002-2908-7360]{Michael R.\ Goad}
\affiliation{School of Physics and Astronomy, University of Leicester, University Road, Leicester, LE1 7RH, UK}

\author[0000-0002-9280-1184]{Diego H.\ Gonz\'{a}lez Buitrago}
\affiliation{Instituto de Astronom\'{\i}a, Universidad Nacional Aut\'{o}noma de M\'{e}xico, Km 103 Carretera Tijuana-Ensenada, 22860 Ensenada B.C., M\'{e}xico}

\author[0000-0002-8990-2101]{Varoujan Gorjian}
\affiliation{Jet Propulsion Laboratory, M/S 169-327, 4800 Oak Grove Drive, Pasadena, CA 91109, USA}

\author[0000-0001-9920-6057]{Catherine J.\ Grier}
\affiliation{Steward Observatory, University of Arizona, 933 North Cherry Avenue, Tucson, AZ 85721, USA}

\author[0000-0002-1763-5825]{Patrick B.\ Hall}
\affiliation{Department of Physics and Astronomy, York University, Toronto, ON M3J 1P3, Canada}

\author[0000-0002-6733-5556]{Juan V.\ Hern\'{a}ndez Santisteban}
\affiliation{SUPA School of Physics and Astronomy, North Haugh, St.~Andrews, KY16~9SS, Scotland, UK}


\author{Chen Hu}
\affiliation{Key Laboratory for Particle Astrophysics, Institute of High Energy Physics, Chinese Academy of Sciences, 19B Yuquan Road, Beijing 100049, People's Republic of China}

\author[0000-0002-1134-4015]{Dragana Ili\'{c}}
\affiliation{Department of Astronomy, Faculty of Mathematics, University of Belgrade, Studentski trg 16,11000 Belgrade, Serbia}
\affiliation{Humboldt Research Fellow, Hamburger Sternwarte, Universit{\"a}t Hamburg, Gojenbergsweg 112, 21029 Hamburg, Germany}

\author[0000-0003-0634-8449]{Michael D.\ Joner}
\affiliation{Department of Physics and Astronomy, N284 ESC, Brigham Young University, Provo, UT, 84602, USA}

\author[0000-0001-5540-2822]{Jelle Kaastra}
\affiliation{SRON Netherlands Institute for Space Research, Niels Bohrweg 4, 2333 CA Leiden, The Netherlands}
\affiliation{Leiden Observatory, Leiden University, PO Box 9513, 2300 RA Leiden, The Netherlands}

\author[0000-0002-9925-534X]{Shai Kaspi}
\affiliation{School of Physics and Astronomy and the Wise Observatory, Tel Aviv University, Tel Aviv 6997801, Israel}

\author[0000-0001-6017-2961]{Christopher S.\ Kochanek}
\affiliation{Department of Astronomy, The Ohio State University, 140 W.\ 18th Ave., Columbus, OH 43210, USA}
\affiliation{Center for Cosmology and AstroParticle Physics, The Ohio State University, 191 West Woodruff Ave., Columbus, OH 43210, USA}

\author[0000-0003-0944-1008]{Kirk T.\ Korista}
\affiliation{Department of Physics, Western Michigan University, 1120 Everett Tower, Kalamazoo, MI 49008-5252, USA}

\author[0000-0001-5139-1978]{Andjelka B. Kova{\v c}evi{\'c}}
\affiliation{University of Belgrade-Faculty of Mathematics, Department of astronomy, Studentski trg 16 Belgrade, Serbia}

\author[0000-0001-8638-3687]{Daniel Kynoch}
\affiliation{Astronomical Institute of the Czech Academy of Sciences, Boční II 1401/1a, CZ-14100 Prague, Czechia}


\author[0000-0001-5841-9179]{Yan-Rong Li}
\affiliation{Key Laboratory for Particle Astrophysics, Institute of High Energy Physics, Chinese Academy of Sciences, 19B Yuquan Road,\\ Beijing 100049, People's Republic of China}


\author[0000-0002-0151-2732]{Ian M. McHardy} 
\affiliation{School of Physics and Astronomy, University of Southampton, Highfield, Southampton SO17 1BJ, UK}

\author[0000-0003-1081-2929]{Jacob N. McLane}
\affiliation{Department of Physics and Astronomy, University of Wyoming, Laramie, WY 82071, USA}

\author[0000-0002-4994-4664]{Missagh Mehdipour}
\affiliation{Space Telescope Science Institute, 3700 San Martin Drive, Baltimore, MD 21218, USA}

\author[0000-0001-8475-8027]{Jake A. Miller}
\affiliation{Department of Physics and Astronomy, Wayne State University, 666 W.\ Hancock St, Detroit, MI, 48201, USA}

\author{Jake Mitchell}
\affiliation{Centre for Extragalactic Astronomy, Department of Physics, Durham University, South Road, Durham DH1 3LE, UK}

\author[0000-0001-5639-5484]{John Montano}
\affiliation{Department of Physics and Astronomy, 4129 Frederick Reines Hall, University of California, Irvine, CA, 92697-4575, USA}
\author[0000-0002-6766-0260]{Hagai Netzer}
\affiliation{School of Physics and Astronomy and the Wise Observatory, Tel Aviv University, Tel Aviv 6997801, Israel}


\author{Christos Panagiotou}
\affiliation{MIT Kavli Institute for Astrophysics and Space Research, Massachusetts Institute of Technology, Cambridge, MA 02139, USA}

\author[0000-0003-1183-1574]{Ethan Partington}
\affiliation{Department of Physics and Astronomy, Wayne State University, 666 W.\ Hancock St, Detroit, MI, 48201, USA}

\author[0000-0003-1435-3053]{Richard W.\ Pogge}
\affiliation{Department of Astronomy, The Ohio State University, 140 W.\ 18th Ave., Columbus, OH 43210, USA}
\affiliation{Center for Cosmology and AstroParticle Physics, The Ohio State University, 191 West Woodruff Ave., Columbus, OH 43210, USA}

\author[0000-0003-2398-7664]{Luka \v{C}.\ Popovi\'{c}}
\affiliation{Astronomical Observatory, Volgina 7, 11060 Belgrade, Serbia}
\affiliation{Department of Astronomy, Faculty of Mathematics, University of Belgrade, Studentski trg 16,11000 Belgrade, Serbia}

\author[0000-0002-6336-5125]{Daniel Proga}
\affiliation{Department of Physics \& Astronomy, 
University of Nevada, Las Vegas 
4505 S.\ Maryland Pkwy, 
Las Vegas, NV, 89154-4002, USA}


\author[0000-0002-5359-9497]{Daniele Rogantini}
\affiliation{MIT Kavli Institute for Astrophysics and Space Research, Massachusetts Institute of Technology, Cambridge, MA 02139, USA}

\author[0000-0003-1772-0023]{Thaisa Storchi-Bergmann}
\affiliation{Departamento de Astronomia - IF, Universidade Federal do Rio Grande do Sul, CP 150501, 91501-970 Porto Alegre, RS, Brazil}


\author[0000-0002-9238-9521]{David Sanmartim}
\affiliation{Carnegie Observatories, Las Campanas Observatory, Casilla 601, La Serena, Chile} 

\author[0000-0003-2445-3891]{Matthew R.\ Siebert}
\affiliation{Department of Astronomy and Astrophysics, University of California, Santa Cruz, CA 92064, USA}

\author[0000-0002-8460-0390]{Tommaso Treu}\thanks{Packard Fellow}
\affiliation{Department of Physics and Astronomy, University of California, Los Angeles, CA 90095, USA}


\author[0000-0001-9191-9837]{Marianne Vestergaard}
\affiliation{Steward Observatory, University of Arizona, 933 North Cherry Avenue, Tucson, AZ 85721, USA}
\affiliation{DARK, The Niels Bohr Institute, University of Copenhagen, Universitetsparken 5, DK-2100 Copenhagen, Denmark}

\author[0000-0001-9449-9268]{Jian-Min Wang}
\affiliation{Key Laboratory for Particle Astrophysics, Institute of High Energy Physics, Chinese Academy of Sciences, 19B Yuquan Road,\\ Beijing 100049, People's Republic of China}
\affiliation{School of Astronomy and Space Sciences, University of Chinese Academy of Sciences, 19A Yuquan Road, Beijing 100049, People's Republic of China}
\affiliation{National Astronomical Observatories of China, 20A Datun Road, Beijing 100020, People's Republic of China}

\author[0000-0003-1810-0889]{Martin J.\ Ward}
\affiliation{Centre for Extragalactic Astronomy, Department of Physics, Durham University, South Road, Durham DH1 3LE, UK}

\author[0000-0002-5205-9472]{Tim Waters}
\affiliation{Department of Physics \& Astronomy, 
University of Nevada, Las Vegas 
4505 S. Maryland Pkwy, 
Las Vegas, NV, 89154-4002, USA}

\author[0000-0002-4645-6578]{Peter R.\ Williams}
\affiliation{Department of Physics and Astronomy, University of California, Los Angeles, CA 90095, USA}


%




\author[0000-0003-0931-0868 ]{Fatima Zaidouni}
\affiliation{MIT Kavli Institute for Astrophysics and Space Research, Massachusetts Institute of Technology, Cambridge, MA 02139, USA}

\author[0000-0001-6966-6925]{Ying Zu}
\affiliation{Department of Astronomy, School of Physics and Astronomy, Shanghai Jiao Tong University, 800 Dongchuan Road, Shanghai, 200240, People's Republic of China}
\affiliation{Shanghai Key Laboratory for Particle Physics and Cosmology, Shanghai Jiao Tong University, Shanghai 200240, People's Republic of China}

\begin{abstract}

We present reverberation mapping measurements for the prominent ultraviolet broad emission lines of the active galactic nucleus \mrk\ using 165 spectra obtained with the Cosmic Origins Spectrograph on the Hubble Space Telescope. Our ultraviolet observations are accompanied by X-ray, optical, and near-infrared observations as part of the AGN Space Telescope and Optical Reverberation Mapping Program 2 (AGN STORM~2). Using the cross-correlation lag analysis method, we find significant correlated variations in the continuum and emission-line light curves. We measure rest-frame delayed responses between the far-ultraviolet continuum at 1180~\AA\ and \Lya\ $\lambda1215$~\AA\ ($10.4_{-1.4}^{+1.6}$~days), \ion{N}{5} $\lambda1240$~\AA\ ($15.5_{-4.8}^{+1.0}$~days), \ion{Si}{4} + \ion{O}{4]} $\lambda1397$~\AA\ ($8.2_{-1.4}^{+1.4}$~days), \ion{C}{4} $\lambda1549$~\AA\ ($11.8_{-2.8}^{+3.0}$~days), and \ion{He}{2} $\lambda1640$~\AA\ ($9.0_{-1.9}^{+4.5}$~days) using segments of the emission-line profile that are unaffected by absorption and blending,  which results in sampling different velocity ranges for each line. However, we find that the emission-line responses to continuum variations are more complex than a simple smoothed, shifted, and scaled version of the continuum light curve. We also measure velocity-resolved lags for the \Lya\, and \ion{C}{4} emission lines. The lag profile in the blue wing of \Lya\ is consistent with virial motion, with longer lags dominating at lower velocities, and shorter lags at higher velocities. The \ion{C}{4} lag profile shows the signature of a thick rotating disk, with the shortest lags in the wings, local peaks at $\pm$ 1500 $\rm km\,s^{-1}$, and a local minimum at line center. The other emission lines are dominated by broad absorption lines and blending with adjacent emission lines. These require detailed models, and will be presented in future work.
\end{abstract}

\section{Introduction}\label{sec:intro}
Supermassive  black  holes (SMBHs)  are  among  the  most extreme  objects  in  the  Universe, and rapidly grow during an active phase of accretion. The symbiotic correlation between SMBH mass (\Mbh) and galaxy bulge properties \citep{Magorrian1998, Gebhardt2000, Ferrarese2000, Gultekin2009, Kormendy2013} implies that black holes are essential ingredients in understanding galaxy evolution. During the active phase of accretion, the galaxy fuels the black hole while outflows in the form of winds and jets may regulate black hole - galaxy coevolution by removing gas from the host galaxy and shutting down \citep{DiMatteo2005, Hopkins2008} or enhancing star formation \citep{Chambers1990, Nesvadba2020}. Despite decades of research, the cosmic evolution of SMBHs and their role in galaxy formation and evolution is still not fully understood. Revealing the nature of gas outflows and their origin near the central black hole are among the foremost requirements for understanding the growth history of SMBHs over cosmic time. 

The masses of nearby SMBHs are generally measured using high spatial resolution observations of gas or stellar dynamics \citep[for a review, see][]{Kormendy2013} with a few exceptions. The mass of Sagittarius A* has been measured using astrometric monitoring of the orbits of individual stars in the central few parsecs of the Milky Way \citep{Ghez2000, Genzel2000}. More recently, the mass of Sagittarius A* and the SMBH in M87 have been measured from radio observations of their black hole ``shadow" \citep{EHT_VI, EHT2022}. There are also interferometric observations of 3C273 \citep{Sturm2018, Wang2020, Li2022} and NGC 3783 \citep{Amorim2021}. Most of these measurements are not possible for more distant galaxies ($>$100 Mpc) even with next-generation facilities. 

Over the past few decades, reverberation mapping (RM; \citealt{Blandford1982, Peterson1993}) has emerged as a powerful technique to study the geometry and kinematics of the gas surrounding SMBHs, and as a tool to estimate the mass of the BHs (for a recent review of multiscale RM, see \citealt{Cackett2021}). Nearly all rapidly accreting SMBHs, observed as broad emission-line active galactic nuclei (AGN), exhibit variability on timescales of weeks to years \citep[e.g.,][]{Collier2001, Peterson2004, Kelly2009,  MacLeod2012}. The gas falling onto the central black hole forms an accretion disk that emits UV photons and ionizes a broad emission-line region (BLR) composed of high-density, high-velocity gas. In its most simplistic implementation, RM can be used to measure the mean light-travel time across the BLR as the time delay (or lag), $\tau$, between the variability in the continuum and the subsequent response of the broad emission-line gas. Assuming that this gas is dominated by the gravitational field of the central black hole, the broad emission-line width (as a line-of-sight gas velocity) is combined with the responsivity-weighted BLR size to obtain an \Mbh\ estimate \citep{Peterson2000}.  

However, the RM \Mbh\ estimate relies on a dimensionless factor $f$, of order unity, that depends on the geometry of the BLR and its orientation relative to the observer's line of sight \citep{Onken2004, Park2012, Woo2010, Grier2013}, which remains poorly understood despite decades of research. Hydrodynamical modeling shows that disk winds can further complicate measurement of $f$ \citep{Kashi2013}.
Some studies of BLR structure support a disk-like BLR \citep{Wills1986, Eracleous1994, Vestergaard2000, Eracleous2003, Strateva2003, Jarvis2006, Gezari2007, Lewis2010, Storchi-Bergmann2017}. 
Other studies suggest that radiation pressure significantly contributes to the BLR dynamics \citep{Marconi2008,Netzer2010}. Further ambiguity in the dynamics of the BLR is caused by the mixed evidence for inflow and outflow processes on these spatial scales (e.g., \citealt{Bentz2009, Denney2009, Barth2011a, Barth2011b, Du2016b, Pei2017}).

A more powerful application of RM is the measurement of velocity-resolved responses of the broad emission lines to continuum flux variations \citep{Bahcall1972, Blandford1982}. Velocity-resolved RM constrains both the distance and kinematics of the broad-line-emitting gas, and enables detailed dynamical modeling of the BLR, but it requires much higher-quality data than velocity-integrated RM \citep{Horne2004}. 
These measurements recover a projection of the BLR structure and kinematics into two observables, the line-of-sight velocity and the emission-line time lag. The distribution of lags can be described as a function of velocity, and it is referred to as the ``velocity-delay map." The broad emission-line flux variations, $\Delta L(V,t)$, are expressed by the convolution of the continuum flux variations $\Delta C(t)$ with the velocity-delay map, $\Psi(V,\tau)$ (also known as the response function) as

\begin{equation}
    \Delta L(V,t) = \int_{0}^{\infty} \Psi(V,\tau) \Delta C(t - \tau) d\tau.
\end{equation}
The primary goal of velocity-resolved RM is to directly constrain the BLR geometry and kinematics either through the inverse problem approach (by reconstructing the velocity-delay map) or through forward modeling. Lag asymmetries between the blue and red wings of a line can be an indication for nonvirial motion of the BLR gas.

Even though these velocity-resolved lags enable simple and qualitative inferences about the BLR gas dynamics, detailed interpretations are dependent on orientation, geometrical complexities and assumptions about the optical depth in the emission line region (see, e.g., \citealt{Welsh1991, Goad2012}). In general, inflows cause shorter lags in the red wing, while outflows lead to shorter lags in the blue wing.  
High-quality velocity-resolved RM campaigns have been successful for a handful of nearby AGN using ground-based observations \citep{Bentz2008, Bentz2009,Denney2009, Bentz2010b, Grier2013, Du2016, Du2018, DeRosa2018, Brotherton2020, Bentz2021, U2022, Bao2022}. These observations have provided results that are consistent with gas in elliptical orbits for some objects, while others indicate either inflowing or outflowing gas trajectories.
Furthermore, the kinematics of the \Hb-emitting region are not uniform across all AGN, with different objects showing a mix of inflowing, stable, and outflowing gas dynamics \citep{Pancoast2014, Williams2018, Li2018, U2022, Villafana2022}. 
Velocity-resolved RM observations for higher ionization ultraviolet (UV) emission lines was only achieved recently through the AGN Space Telescope Optical Reverberation Mapping (AGN STORM) program.

The first AGN STORM campaign was a pioneering multiwavelength reverberation mapping program, targeting NGC\,5548 over 6 months with daily spectroscopic observations using the Hubble Space Telescope (\hst) Cosmic Origins Spectrograph (COS) in 2014 \citep{DeRosa2015}. The \hst\ observations were also supported by coordinated observations from Swift using the XRT and UVOT instruments \citep{Edelson2015}, intensive ground-based photometric and spectroscopic monitoring \citep{Fausnaugh2016, Pei2017}, and four Chandra X-ray observations distributed over the duration of the campaign \citep{Mathur2017}. Key results from the AGN STORM campaign include the following.
\begin{itemize}
    \item Identification of a stratified BLR with mean broad emission-line delays spanning 2.5 -- 7 days with respect to the $\lambda 1367$~\AA\ continuum, depending on the emission line \citep{DeRosa2015}. When placed in the context of previous intensive monitoring campaigns of this source (e.g., \citealp{Clavel1991, Peterson1991, Denney2009}), the continuum fluctuation timescales were substantially shorter even though continuum fluctuation amplitudes and average luminosities were similar. The shorter emission-line delays and smaller variability amplitudes were in part due to the short timescale continuum fluctuations.
    \item The relative time lags between the variations in the inner/shorter-wavelengths and outer/longer-wavelengths of the disk were measured through continuum reverberation from the UV through the near-infrared (NIR) wavelength region. These measurements are broadly consistent with the temperature gradient expected from the foundational ``thin-disk" model of \citet{SS1973}, but suggest a disk that is three times larger \citep{McHardy2014, Edelson2015, Fausnaugh2016}. However, these measurements may be affected by contributions from diffuse continuum emission from the BLR clouds \citep{Korista2001, Korista2019, Chelouche2019}. 
    \item The appearance of anomalous BLR behavior (the so-called ``BLR Holiday") started midway through the campaign \citep{Goad2016, Pei2017, Kriss2019, Dehghanian2019a}. This was a ~40-day period where the emission line light curve was uncorrelated with the continuum. Anomalous behavior was also later identified in the continuum bands \citep{Goad2019}, further supporting contamination of the UV-optical continuum by the diffuse continuum from the BLR.
    \item Recovery of the most detailed velocity-delay maps ever obtained for the prominent emission lines revealed kinematics that are dominated by near-circular Keplerian motion, and a weaker response from the far side of a somewhat flattened BLR geometry \citep{Horne2021}.
    \item The black hole mass derived from a reverberation analysis using a mean time delay, a characteristic emission line velocity dispersion, and a scale factor of $\langle f\rangle \approx 5$ is in good agreement given the uncertainties with the mass derived from dynamical modeling utilizing velocity-delay mapping information \citep{Williams2020}. The masses inferred from independent emission lines are mutually consistent, even though the geometry and kinematics of each line-emitting region are different.
\end{itemize}
The present work describes the results of a similar RM campaign on \mrk, a large monitoring program built around 165 epochs of \hst\ observations\footnote{HST-GO-16196; \citet{Peterson2020}}. Hereafter, we refer to the campaign on NGC\,5548 as AGN STORM~1 and the current program on \mrk\ as AGN STORM~2.

The target for AGN STORM~2, \mrk, was carefully selected because it showed no history of strong UV absorption and/or X-ray obscuration. Therefore, it was less likely to be obscured, which was a major source of complications in AGN STORM~1 \citep{Goad2016, Kriss2019, Dehghanian2019a}. Historically, \mrk\ has exhibited weak, variable intrinsic narrow absorption features, which can be used to probe the unobservable UV ionizing continuum along the line of sight \citep{Kriss2019, Dehghanian2019a, Dehghanian2019b}. The \mrk\ SMBH has similar mass to that of NGC\,5548 ($M_{\rm BH} \approx 3.85\times 10^7 M_{\odot}$); however, its higher Eddington ratio ($L/L_{\rm Edd} \approx 0.2$ compared to $L/L_{\rm Edd} \approx 0.03$ for NGC\,5548) probes a new region of parameter space compared to NGC\,5548. Furthermore, the higher redshift of \mrk\ ($z = 0.03146$) ensures that the \Lya\ line is not affected by geocoronal emission and Galactic absorption features. Also, \mrk\ is located near the ecliptic pole, making it continuously observable.

Even though \mrk\ was carefully selected to be free of strong UV absorption lines, the AGN STORM~2 observations reveal the presence of new broad and narrow absorption lines, which may indicate the appearance of a new dust-free wind located at the inner BLR that partially obscures the line of sight to the SMBH \citep{Kara2021}. Furthermore, coordinated X-ray observations with XMM-Newton and NICER show that the X-rays are significantly obscured compared to earlier observations (see \citealt{Miller2021} for an independent analysis using NuSTAR data). \citet{Kara2021} provide a detailed discussion of these early results, describing the HST program from the first 97 days of the AGN STORM~2 campaign.

This paper is the second in a series of the AGN STORM~2 campaign. We discuss the calibration and improvements of the spectra from the \hst\ Cosmic Origins Spectrograph (COS; \citealt{Green2012}) for the full duration of the AGN STORM~2 campaign. In \autoref{sec:Obs} we discuss the details of the observations. Section \autoref{sec:data} describes our custom reduction pipeline. We present the analysis of spectra, flux, and light-curve measurements in \autoref{sec:Analysis}. In \autoref{sec:timeseries} we describe the light curves and their cross correlations. In \autoref{sec:discussion} we discuss the velocity-binned reverberation lags for the major UV emission lines. Throughout this work, we adopt a $\Lambda$CDM cosmology with $\Omega_{\Lambda}$ = 0.7, $\Omega_M$ = 0.3, and $H_0$ = 70 km $\mathrm{s^{-1}\,Mpc^{-1}}$.

\section{Observations} \label{sec:Obs}
\subsection{Program Design}\label{sec:design}
The ultimate goal of the AGN STORM~2 campaign is to understand the origins of gas flows in the nuclear region, including the location and dynamics of the BLR. This is achieved through the construction of velocity-delay maps or through forward modelling using high-quality RM observations. To obtain such a dataset, several considerations are required in the design of the observations, as follows.
\begin{itemize}
    \item Observing cadence: A velocity-delay map with time resolution $\Delta t$ requires data with time sampling on scales $\Delta t/2$ or shorter.
    \item Signal-to-noise ratio ($\rm{S/N}$) and spectroscopic calibration: Continuum and emission-line variations are small on short timescales and therefore high $\rm{S/N}$ is required to successfully detect them. In practice, this means $\rm{S/N}\ge100$ in the integrated continuum windows (measured over several resolution elements), and local flux calibrations precise and stable to $\sim$ 1-2\%.
    \item Spectral resolution: The line-of-sight (LOS) velocity resolution should be sufficient to resolve structures in the emission and absorption line profiles on scales of a few hundred $\rm{km\,s^{-1}}$.
    \item Duration: The key to obtaining a clear reverberation signal is structure in the light curves. The temporal coverage must be long enough to both have a significant probability of including such features and track the subsequent response of the BLR.
\end{itemize}
Previously, \mrk\ has been the target of ground-based RM campaigns \citep{Peterson1998, Denney2010}. These campaigns show an optical continuum root-mean-square (RMS) variability amplitude of 5-14\% on timescales of weeks to months. The AGN UV variability is expected to be twice as large \citep{Korista1995,Marshall1997, MacLeod2012}. The \Hb\ BLR size in \mrk\ (14-34 light days) is larger compared to the BLR size in NGC\,5548, which was observed during the AGN STORM~1 campaign. For a historically variable target such as \mrk, an every-other-day observing cadence over 365 days provides the necessary duration while also capturing UV-emitting-region variability (compared to daily cadence over a six-month duration in the AGN STORM~1 campaign).


\subsection{HST Observations}\label{sec:obs}
\mrk\ was observed with \hst\ COS over 165 epochs with a median cadence of 2 days from 2020 November 24 to 2022 February 24. 
These data are available from MAST: \dataset[doi:10.17909/10sp-zt74]{https://doi.org/10.17909/10sp-zt74}.
We obtained high-resolution COS spectra in single-orbit visits using the G130M (10,000 $< R <$ 15,000) and G160M (13,000 $< R <$ 24,000) gratings to obtain far-UV spectra covering 1070 -- 1750~$\mathrm{\AA}$. For each visit, we obtained four 60 s exposures with the G130M grating centered at 1222~$\mathrm{\AA}$, two exposures of 175 s and 180 s using the G160M grating centered at 1533~$\mathrm{\AA}$, and two 195 s exposures with G160M at 1577~$\mathrm{\AA}$. These exposure times ensure a $\mathrm{S/N}\approx10$ per resolution element in the continuum at 1180~$\mathrm{\AA}$ and 1502 ~$\mathrm{\AA}$.

For COS far-UV (FUV) observations, moving the spectra slightly in the dispersion direction minimizes the effects of small-scale fixed-pattern noise, detector artifacts, and grid-wire shadows in the detector, and improves the S$/$N of the coadded spectra \citep{Dashtamirova2020}. Furthermore, prolonged exposure of the same detector area to bright targets causes it to become less efficient at photon-to-electron conversion over time (a phenomenon also known as ``gain-sag"). While the bright \Lya\ airglow falls in the detector gap for the G130M/1222 setting, repeated illumination by the AGN emission lines highly affects the detector sensitivity.
To mitigate the issue, we ensured that the location of the spectrum on the detector was periodically moved by alternating the G130M/1222 exposures between four grating offset focal-plane positions (FP-POS=1$-$4), the G160M/1533 exposures between two FP-POS configurations (FP-POS=1,2), and the G160M/1577 exposures between the remaining 2 FP-POS settings (FP-POS=3,4). 

We also used 13 additional orbits to observe a flux standard star (WD0308$-$565) periodically throughout the campaign, in order to obtain high-precision flux calibrations on small spectral scales, as outlined in section~\ref{sec:design}. We observed WD0308$-$565 approximately once per month, using exactly the same settings as the RM program.

In addition to the COS spectra, we dedicated 6 additional orbits to obtain low-resolution spectra using the STIS/NUV-MAMA and the STIS CCD to capture full spectra from the far-UV to the near-IR. We used gratings G230L, G430L, and G750L in 2-orbit visits paired with COS observations to cover the range 940--10,200 $\rm \AA$. These visits were evenly distributed throughout the campaign and they were coordinated with XMM-Newton observations, as described by \citet{Kara2021}. The full-spectrum snapshots plus the X-ray spectra provide information on the spectral energy distribution of \mrk. 
The close agreement (generally $<$5\%) between the 1600--1700 \AA\ fluxes in the STIS
observations (which used a 0\farcs2 slit) and the COS observations in the same visit (2\farcs5  circular aperture) show that any host galaxy contribution to the COS continuum flux is negligible.

The results of the combined COS and STIS spectra will be released in a separate paper.

Out of 165 epochs, eight visits suffered a failed guide-star acquisition or a guide-star loss-of-lock, leading to empty exposures. Furthermore, \hst\ suffered two safe-mode incidents resulting in large gaps in our campaign. The first \hst\ safing incident occurred on 2021 June 12 and lasted for 36 days. The second \hst\ safing occurred on 2021 October 25, at the beginning of an expected 16-day observing gap due to no guide-star availability, and lasted for 51 days. 
Our original observing plan anticipated that 10\% of visits might fail due to HST gyro issues, which is similar to the final rate of failed visits (11\%).
All of the lost visits were compensated and added to the end of the campaign at nominal cadence. As a result, the observing campaign ended on 2022 February 24, instead of the expected end date of November 2021. 

\section{Data Reduction} \label{sec:data}
We used the \texttt{CalCOS} pipeline v3.4.1 for the bulk of our data processing. In order to successfully perform spectral modeling and velocity-resolved RM, we need the spectra to be locally\footnote{single-pixel level} accurate, precise, and stable. The standard COS reduction pipeline guarantees a flux accuracy at the 5\% level, a global precision of 2\%, and a wavelength accuracy of better than $10~\rm km~s^{-1}$ for G130M/G160M gratings. While the absolute flux accuracy we require is achieved with the automated reduction pipeline, we need the flux measurements to be locally precise to $<$2\%. Therefore, after testing the repeatability and precision of the wavelength solution, we refined the existing
flux calibration reference files and applied a post-\texttt{CalCOS} pipeline
to further process the data, following a similar approach to the one described by \citet{DeRosa2015} in AGN STORM~1. 
The main areas of improvement
include refinements in the sensitivity function and the time-dependent sensitivity (TDS) correction, as outlined below.
These reprocessed data and the various forms of combined spectra described below
are available as a high-level science product in the
\href{http://archive.stsci.edu}{{\it Mikulski Archive for Space Telescopes} (MAST)} as the data set identified by
\dataset[https://doi.org/10.17909/n734-k698]{https://doi.org/10.17909/n734-k698}. The reference files used in the custom pipeline are also avaiable through the same DOI.

\subsection{Dispersion Solution}\label{sec:disp}
To verify the precision of the dispersion solution for each separate grating setting and each separate observation, we compare the registration of all the campaign spectra. In absolute terms, we also compare the mean positions of selected interstellar medium (ISM) lines to direct line-of-sight 
observations of Galactic \ion{H}{1} 21-cm emission from \cite{Murphy96}.
The \ion{H}{1} emission is slightly blueshifted by $- 16~\rm km~s^{-1}$ in the heliocentric frame, with additional components extending blueward to $- 80~\rm km~s^{-1}$.
The ISM lines in the COS spectrum have a similar structure, with the deepest point in the troughs agreeing with the \ion{H}{1} to better than
$10~\rm km~s^{-1}$.

\subsection{Sensitivity Function and TDS}\label{sec:tds}
To obtain a local precision of the flux calibrations at  better than 2\%, we used approximately monthly observations of the \texttt{CALSpec} calibration standard WD0308$-$565. The standard-star observations were obtained using exactly the same gratings, central wavelength (CENWAVE) settings, and focal-plane positions (FP-POS) as the science exposures. We estimate the quality of the flux calibration by inspection of the fractional residuals 
\begin{equation}\label{eq:tds}
f_{\rm res} = \frac{f_{\rm WD, Obs}-f_{\rm Model}}{f_{\rm Model}},
\end{equation}
of the calibrated standard-star spectra and their respective \texttt{CalSPEC} stellar model (we used the latest model available for WD0308$-$565), where both $f_{\rm WD,obs}$ and $f_{\rm Model}$ are binned over 1 \AA \  using a boxcar filter in order to increase the S/N per spectral element. By analyzing these calibration data, we verified that the residuals at each epoch, as well as their variations through the campaign, were higher than the AGN STORM\,2 requirements and thus need further calibrations.

The COS flux calibration is done in two steps: (a) characterization of the time evolution of the sensitivity through the TDS correction, and (b) derivation of static sensitivity functions, obtained once for each lifetime position.

The COS reference TDS model is obtained from approximately bimonthly observations of the standard stars WD0308$-$565 and GD71 as part of the yearly COS TDS monitoring program. Observations are obtained for a selected subset of CENWAVEs, including the bluest and reddest CENWAVE for each grating at one fixed FP-POS (FP-POS=3). The COS monitoring program covers all of the CENWAVEs used in our observations (1222~\AA, 1533~\AA, 1577~\AA). However, for some of the CENWAVEs (e.g., G160M/1577), the FUVA and FUVB sensitivity functions are obtained from observations of different \texttt{CalSPEC} standards, which can affect the resulting spectral shape due to intrinsic systematic uncertainties in the stellar models. To improve the COS monitoring time cadence in our custom reduction, we utilized the existing COS monitoring program data and additionally added our observations of WD0308$-$565 at FP-POS=3 only for G130M/1222 and G160M/1577 and FP-POS=1 and 2 for G160M/1533 because no COS monitoring data is available for FUVB with WD0308$-$565 for that CENWAVE. We then refined the TDS model by increasing the number of time intervals over which the TDS trends are computed and by redefining the wavelength ranges used in the analysis.

\begin{figure*}
\centering
\includegraphics[width=\textwidth]{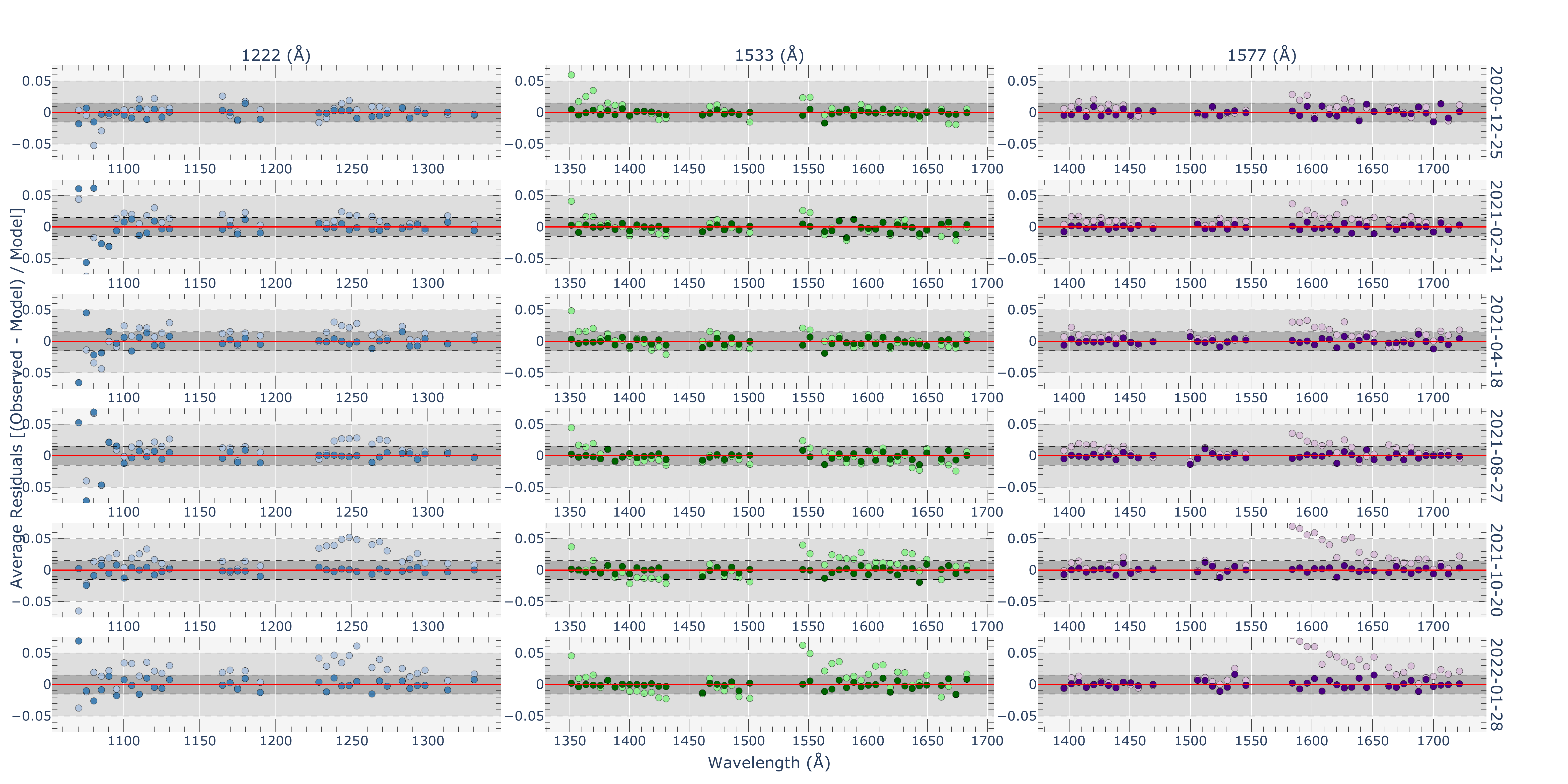}
\caption{Flux residuals $f_{\rm res}$ versus wavelength for each of the three central wavelengths used to observe \mrk: 1222 \AA\ (left, blue), 1533 \AA\ (middle, green), and 1577 \AA\ (right, purple). Six of the twelve white-dwarf epochs are shown. In each panel, the lighter circles represent the residuals obtained with \texttt{CalCOS}, while the darker circles are obtained with the AGN STORM\,2 updated data reduction. The light gray band and dark gray band show $\pm5\%$ and $\pm1.5\%$ residuals, respectively.}
\label{fig_cal}
\end{figure*}

In the standard \texttt{CalCOS} reduction, one sensitivity function is used per lifetime position and any time variations are resolved in the TDS correction. However, we see variations in the fractional residuals (Equation~\ref{eq:tds}) over time that could not be accounted for with any additional TDS corrections. We therefore believe that the variations may come from uncharacterized changes in the standard flux or flats, so, to improve the flux calibrations, we obtained 12 new sensitivity functions using our monthly observations of WD0308$-$565. The \mrk\ data were then reduced using the sensitivity function that was truncated in time to the science observation date and the updated TDS correction, where any observation before that time was reduced using the sensitivity function from the previous month. The improved residuals of our calibrations are illustrated in Figure~\ref{fig_cal}.
 
We use the standard deviation of the residuals for each grating as an estimate of the fractional precision error $\delta P$ to be associated with each epoch. $\delta P$ is a proxy for the stability of the flux calibration at a given time. Since our final G160M spectra are obtained from the combination of two CENWAVEs, we conservatively define the fractional precision error for the G160M grating as the maximum fractional uncertainty computed for the 1533~\AA\ and 1577~\AA\ CENWAVEs, respectively. Table~\ref{tab:precision} lists the final fractional precision as a function of time for G130M and G160M.

\begin{table}
\caption{Fractional Precision Error $\delta P$}\label{tab:precision}
\begin{center}
  \begin{tabular}{cccc}
    \hline
    \hline
    HJD Start & HJD End & G130M & G160M \\
    {(Days)} & {(Days)} & {\%} & {\%}\\
    \hline
    9178 & 9206 & 0.8 & 0.6 \\
    9207 & 9229 & 0.8 & 0.8 \\
    9230 & 9264 & 0.8 & 0.7 \\
    9265 & 9296 & 0.9 & 0.8 \\
    9297 & 9320 & 0.5 & 0.8 \\
    9321 & 9411 & 0.8 & 0.6 \\
    9412 & 9451 & 1.3 & 0.7 \\
    9452 & 9476 & 0.9 & 0.7 \\
    9477 & 9505 & 0.9 & 0.7 \\
    9506 & 9547 & 0.8 & 0.7 \\
    9548 & 9605 & 1.6 & 0.7 \\
    9606 & 9633 & 0.9 & 0.7 \\ 
    \hline
 \end{tabular}
\end{center}
\end{table}

\subsection{Spectral Combination and Error Arrays }\label{sec:FinalSpec}
The final data product consists of one combined spectrum per grating per epoch. 
The final spectra are binned by 4 pixels in order to increase the S/N per spectral element of the AGN continuum. This binning still results in two binned macropixels per COS resolution element. For the G160M grating, the spectra from the G160M/1533 and G160M/1577 settings are merged together and resampled over a 4-pixel interval, 
using a python implementation of the \texttt{iraf/splice} algorithm. Each pixel is weighted by the DQ\_WGT array in the COS data files to correctly account for masked bad pixels. Finally, we modify the flux error obtained by propagating the \texttt{CalCOS} statistical uncertainty to include a correction for low number statistics following \citet{Gehrels1986}; the correction is applied in counts space. This is particularly important for regions of the combined spectra characterized by low counts (e.g., the blue end of G130M, the red end of G160M, and the troughs of the deepest absorption lines).

\section{Data Analysis} \label{sec:Analysis}
To measure spectral variations and time-series quantities, we follow steps similar to those of \citet{DeRosa2015}. We first compute the mean and root-mean square (RMS) spectra from the 165 epochs of G130M and G160M observations. The RMS spectra are used to isolate the variable spectral components. This clears the way for defining integration limits for the continuum and line light curves as outlined below (see Table~\ref{tab:integration}). 

\subsection{Mean and RMS Spectra}\label{sec:meanrms_spec}
\begin{figure*}
\centering
\includegraphics[width=0.9\textwidth]{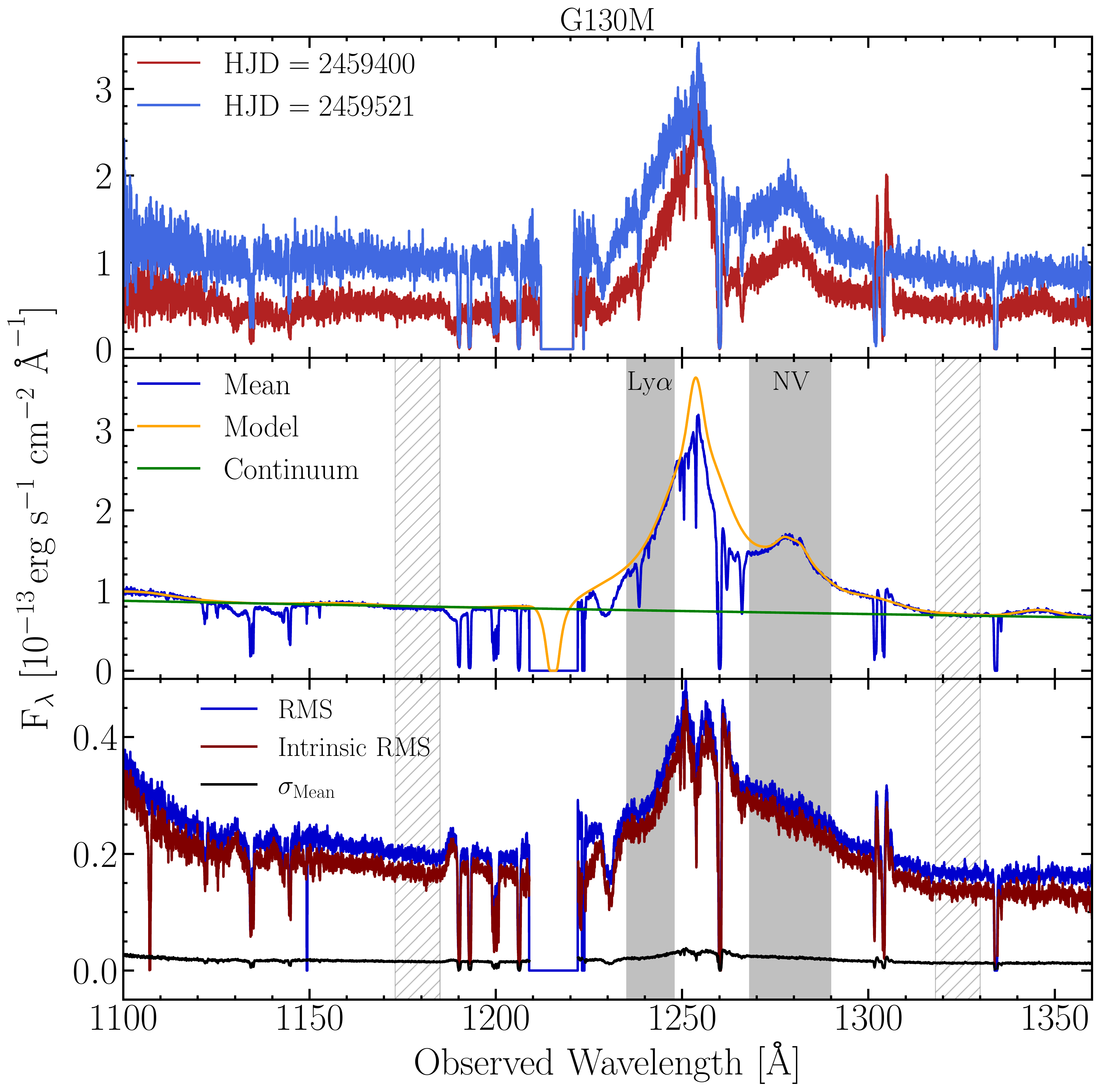}
\caption{The top panel illustrates spectra from the epochs of lowest flux with strong absorption (red), and highest flux with weak absorption (blue). The red spectrum corresponds to HJD = 2459400, and the blue spectrum was obtained at HJD = 2459521. The middle and bottom panels show the mean and RMS spectra, respectively, for the 1100--1360 $\mathrm{\AA}$ range of the G130M grating. The solid gray shaded regions show the integration windows for \Lya\ and \ion{N}{5}, and the hashed gray regions illustrate the shortward and longward continuum windows used for continuum subtraction (see Table \ref{tab:integration}). The mean spectrum is computed from Equation \ref{eq:mean_spec}. The empirical model of the intrinsic emission lines and continuum are superposed on the mean spectrum. In the bottom panel, the RMS spectrum as defined in Equation \ref{eq:rms_spec} is shown in blue and the intrinsic RMS spectrum is illustrated in red. The black line shows the total uncertainty in the mean, which is computed from the square root of the quadrature sum of statistical uncertainties (Equation \ref{eq:mean_stat_err}) and the uncertainties in precision of flux calibration (see Table~\ref{tab:precision}) for the G130M grating.}
\label{fig2_g130}
\end{figure*}

\begin{figure*}
\centering
\includegraphics[width=0.9\textwidth]{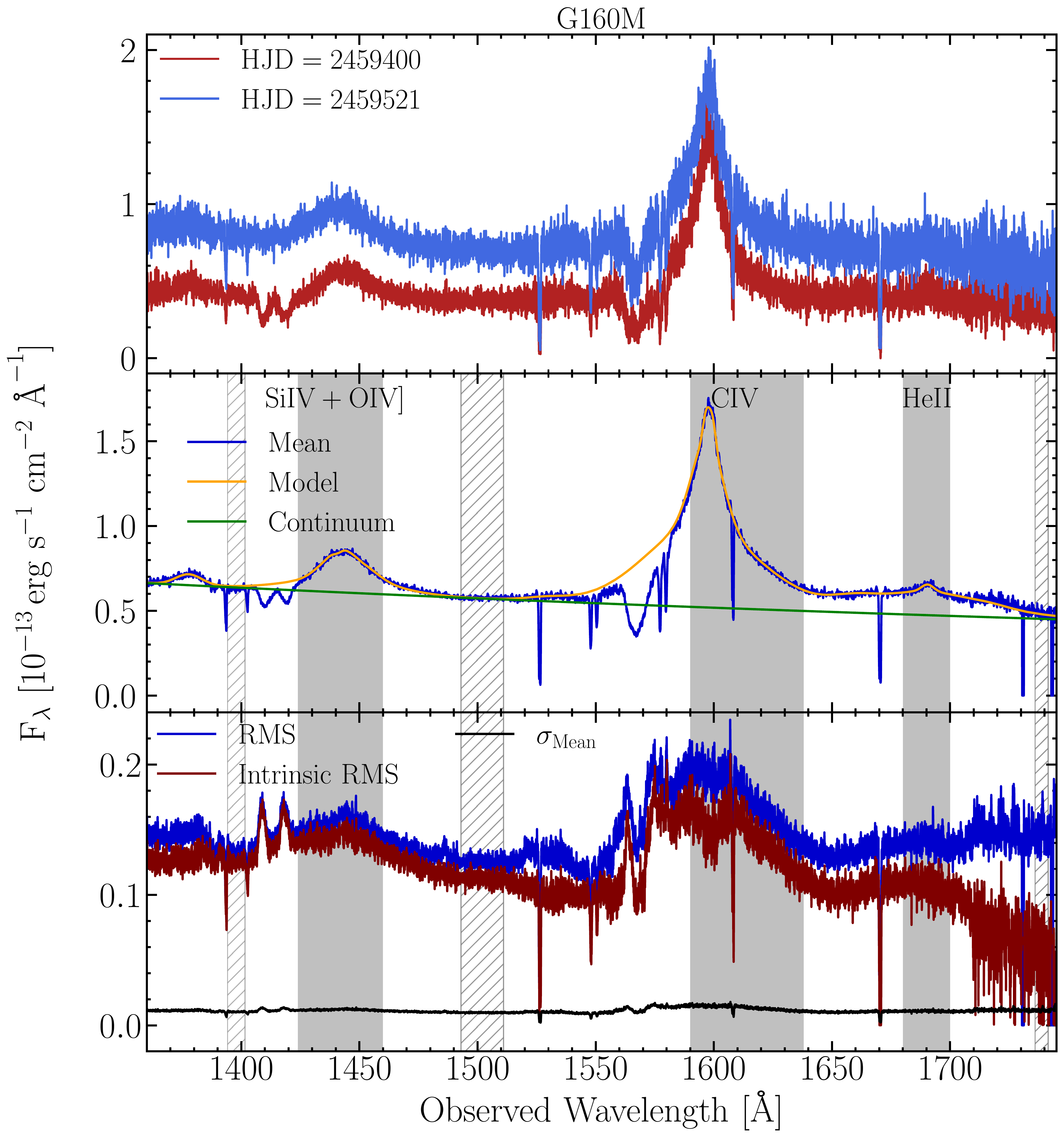}
\caption{The top panel illustrates spectra from the epochs of lowest flux with strong absorption (red), and highest flux with weak absorption (blue). The red spectrum corresponds to HJD = 2459400, and the blue spectrum was obtained at HJD = 2459521. The middle and bottom panels show the mean and RMS spectra, respectively, for the 1360--1700 $\mathrm{\AA}$ range of the G160M grating. The solid gray shaded regions show the integration windows for the \ion{Si}{4} + \ion{O}{4]} blend, \ion{C}{4}, and \ion{He}{2} emission lines, and the hashed gray regions illustrate the continuum windows used for each emission line continuum subtraction (see Table \ref{tab:integration}). The mean spectrum is computed from Equation \ref{eq:mean_spec}. The empirical model of the intrinsic emission lines and continuum are superposed on the mean spectrum. In the bottom panel, the RMS spectrum as defined in Equation \ref{eq:rms_spec} is in blue. The intrinsic RMS spectrum is in the red. The black line shows the total uncertainty on the mean, which is computed from the square root of the quadrature sum of statistical uncertainties (Equation \ref{eq:mean_stat_err}) and the uncertainty in precision of flux calibration (see Table~\ref{tab:precision}) for the G160M grating.}
\label{fig3_g160}
\end{figure*}

The mean spectrum for the set of G130M and G160M spectra is defined as
\begin{equation}\label{eq:mean_spec}
\bar{F}(\lambda) = \frac{1}{N} \sum_{i=1}^{N}\ F_i(\lambda)
\end{equation}
where $F_i$ is the $i$th spectrum of the series of $N$ = 165 spectra, and the RMS spectrum is defined as

\begin{equation}\label{eq:rms_spec}
S(\lambda) = \Big(\,\frac{1}{N-1} \sum_{i=1}^{N}\big(F_i{(\lambda)} - \bar{F}(\lambda)\big)^2\Big)^{1/2}.
\vspace{0.05cm}
\end{equation}
The mean and RMS spectra for the G130M and G160M settings are shown in Figures \ref{fig2_g130} and \ref{fig3_g160}, respectively. The statistical uncertainty in the mean spectra is
\begin{equation}\label{eq:mean_stat_err}
\sigma_{\bar{F}}(\lambda) = \frac{1}{N}\Big[\sum_{i=1}^{N} \sigma_{F_i}^2(\lambda) \Big]^{1/2},
\end{equation}
where $\sigma_{F_i}$ is the error spectrum of the $i$th spectrum in the series. We estimate the total uncertainty of the mean spectrum as the quadratic sum of the statistical uncertainty and fractional precision $\delta_p$.

The RMS spectrum computed from Equation \ref{eq:rms_spec} and illustrated in blue in Figures \ref{fig2_g130} and \ref{fig3_g160} includes the intrinsic variability, $\sigma_0$, as well as the variance due to noise (\citealt{Park2012, Barth2015}). We follow \citet{DeRosa2015} and model the distribution of the residuals of each pixel about the mean to estimate $\sigma_0$ using a maximum-likelihood estimator for the optimal average.
The estimated intrinsic RMS spectrum, $\sigma_0$, obtained from the 165 \hst\ COS spectra is shown in red (see Figures \ref{fig2_g130} and \ref{fig3_g160}).

 The intrinsic RMS spectrum reveals several significant features of the spectral variability observed in the AGN STORM~2 campaign. The broad emission lines themselves, the primary motivation for the campaign, show strong variability across their entire profiles. However, the variability amplitude varies across the different line components. For example, in the \ion{He}{2} line, the majority of the variability is in a very broad component. There is also a strong rise in variability at the blue end of the spectrum, which is due to the red wing of the broad \ion{O}{6} $\lambda1033$~\AA\ emission line. Furthermore, we note that the \ion{C}{4} RMS profile is broader compared to its mean profile, which is similar to other studies of the \ion{C}{4} emission line that could be indicative of a non-variable core component \citep{Korista1995, Denney2012}, and has raised concerns over the suitability of \ion{C}{4} for SE \Mbh\ estimates.

The broad absorption troughs also produce distinctive signatures. The saturated portions of the troughs typically show less variability than surrounding portions of the spectrum. This is particularly noticeable in \Lya, \ion{N}{5}, and the centers of the \ion{C}{4} doublet. Both transitions in the \ion{N}{5} doublet display prominent troughs at the peak of \Lya\ at 1254~\AA\ and on its red wing at 1259~\AA. Note that these saturated troughs have shapes quite similar to the saturated broad \Lya\ trough at 1230~\AA. The weaker, potentially more optically thin portions of the obscuring outflow, however, show greater variability, perhaps indicating a more immediate response to continuum variations. This is evident in the \ion{C}{3}*~$\lambda 1176$ trough at 1189~\AA, the troughs in the \ion{Si}{4} doublet, and the red and blue wings of the \ion{C}{4} trough.

We take the presence of these features into account when we evaluate the windows for extracting continuum and emission-line fluxes in the next section.

\subsection{Integrated Light Curves}\label{sec:lightcurves}
We extract the light curves for the continuum and the prominent UV emission lines listed in Table~\ref{tab:integration} by performing the spectral flux measurement in the observed frame. We have not corrected the spectra for Galactic extinction in order to facilitate the cleanest comparison with other measurements to be reported elsewhere in this series of papers (e.g., broadband photometry).

There are bad pixels throughout the spectrum, and their location and severity change with time and instrument settings. To prevent the introduction of artificial variations in the relative-flux estimates, bad pixels are masked throughout the dataset. This means that if a pixel is bad in any of the visits, the pixel is masked out in all of the 165 spectra. We have also masked out Galactic \Lya\ absorption in the range 1209--1222 $\mathrm{\AA}$. The presence of intrinsic broad and narrow absorption features that appeared during the AGN STORM~2 campaign complicates the analysis of the spectra. Final integration ranges (listed in Table \ref{tab:integration}) were chosen using the empirical model of the spectra and the variability characteristics of the RMS spectrum.
Our modeling is similar to that used by \citet{Kriss2019} for the AGN STORM~1 campaign, which includes the broad absorption features associated with all permitted transitions in the spectrum. \citet{Kara2021} discuss the modeling approach adopted here in detail.

The heuristic spectral model for \mrk\ is shown in the upper panels of Figures \ref{fig2_g130} and \ref{fig3_g160}, superposed on the observed mean spectra. 
We chose continuum ranges that are least affected by absorption-line contamination and broad emission-line wings. Some of the emission lines (e.g., \Lya\ -- \ion{N}{5} and \ion{C}{4} -- \ion{He}{2}) have overlapping wings. In these cases, the boundary wavelength corresponds to the wavelength at which the fluxes of the two lines are comparable. We do not mask absorption lines at this stage in our analysis, so the regions chosen to measure the emission-line fluxes do not account for all flux in each of the lines. We report the total fraction that is included in our analysis for each line flux measurement in Table~\ref{tab:integration}, where we integrate the flux in the mean model spectrum over the chosen wavelength regions. Detailed analysis of the overlapping emission lines requires in-depth modeling of individual emission and absorption lines as performed by \citet{Kriss2019}, which will be provided in a separate paper.

We use the identified integration limits in Table \ref{tab:integration} to measure the continuum fluxes using the weighted mean of the flux density in the integration region with weights defined as the inverse of the variance (square root of the propagated error). 

\begin{figure*}[tt]
\centering
\includegraphics[width=0.9\textwidth]{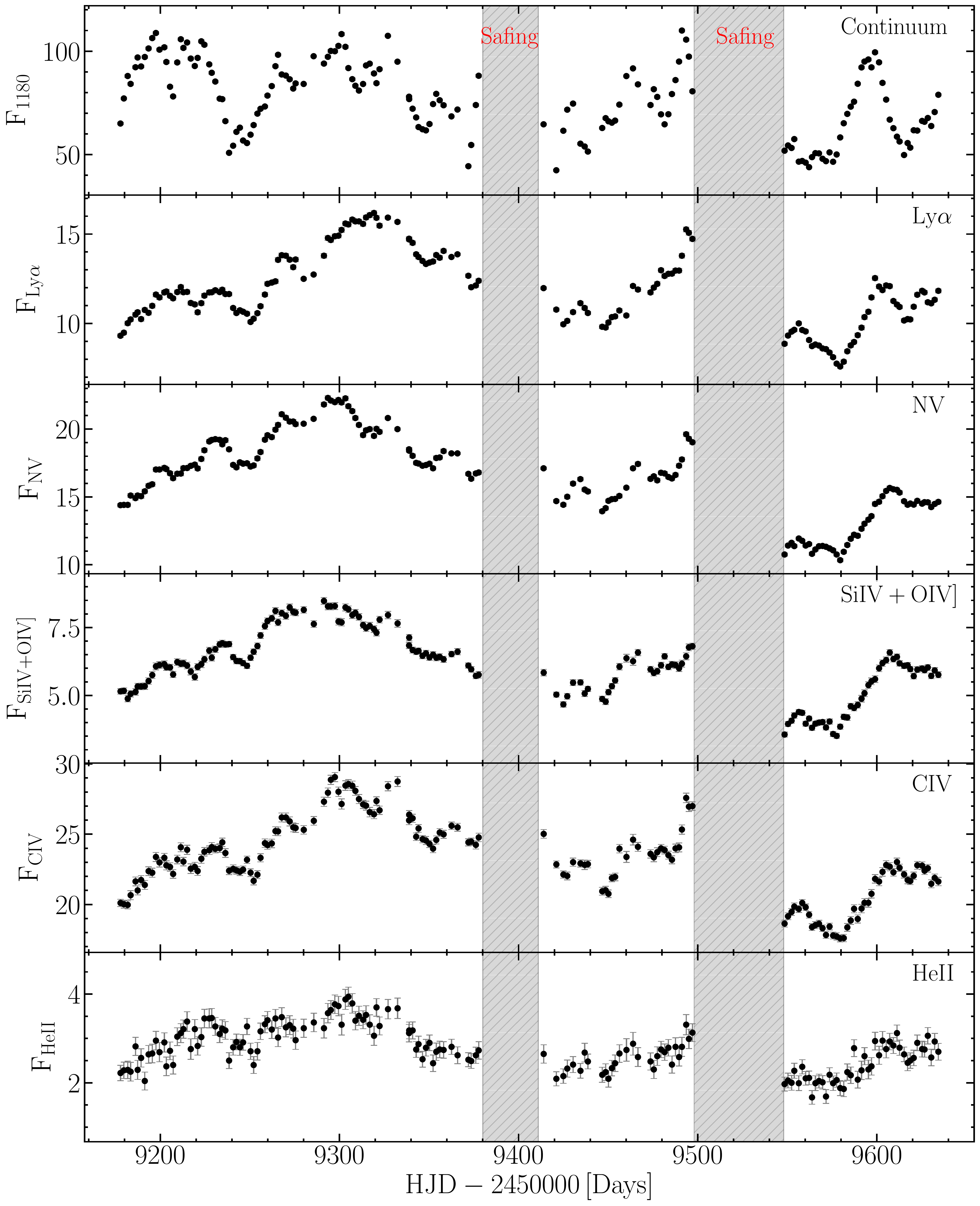}
 \caption{Integrated continuum light curve at 1180 $\mathrm{\AA}$ (top panel) and emission-line light curves for \Lya, \ion{N}{5}, \ion{Si}{4} + \ion{O}{4]} blend, \ion{C}{4}, and \ion{He}{2}, respectively, in panels below the top panel. The \Lya\ light curve is the integrated flux in the blue wing of the emission line over the region not contaminated by absorption, and thus only includes 21\% of the total emission line profile (see Table~\ref{tab:integration}). All the light curves are computed from the observed-frame spectra. The continuum light curve is in units of $10^{-15}\, \mathrm{erg\, s^{-1}\, cm^{-2} \AA^{-1}}$ and the line fluxes are in units of $10^{-13}\, \mathrm{erg\, s^{-1}\, cm^{-2}}$. The hashed gray regions display the two major observation gaps during our campaign, which were the result of \hst\ safing events.
 }
\label{fig4_lcs}
\end{figure*}

Emission-line fluxes are measured as the numerical integral of the emission flux above a local continuum defined by a linear fit to the continuum flux in the continuum region adjacent to each emission line (see Table~\ref{tab:integration}). The linear local continuum is then subtracted from the emission component. The line flux is numerically integrated over the integration limits given in Table \ref{tab:integration} using Simpson's method and excluding masked pixels. Statistical errors are computed numerically by creating $N_{\mathrm{sample}}$ = 5000 realizations of the line flux and the underlying linear continuum \citep[see][]{DeRosa2015}. The flux, $F_{\lambda}$, in each realization is randomly generated from a Gaussian distribution having a mean equal to the flux of the spectral element and the width $\sigma$ equal to the statistical error on the flux. For the linear continuum, we generate $N_{\mathrm{sample}}$ models using the best-fit values and covariance matrix of the linear fit to the data. For each realization, a line-flux estimate is then obtained by subtracting the linear continuum and then performing the numerical integration of the residuals. The $1\sigma$ confidence levels are finally obtained from the distribution of the $N_{\mathrm{sample}}$ line fluxes. When the uncertainties are asymmetric, we adopt the larger of the two uncertainties as the statistical uncertainty associated with the integrated flux. Figure~\ref{fig4_lcs} shows the final full-campaign continuum and line light curves. We report the light-curve statistics in Table~\ref{tab:stats}. Tables~\ref{tab:lcs_con} and \ref{tab:lcs_line} show a sample of the custom-calibrated light curves, which are available in machine-readable form in the online version of this paper.

The light curves in Figure \ref{fig4_lcs} dramatically illustrate the high quality of the AGN STORM~2 dataset. The continuum variations are strong, with at least ten strong peaks more than 20\% higher than the adjacent minima, and several with a factor $\sim 2$ contrast. Smaller variations on several-day timescales are readily visible, and they are well sampled by the two-day cadence and easily discernible given the small uncertainties ($<$1.5\%) in our processed data. These features make the differing characters of the continuum light curve and the emission-line light curves obvious. Early in the campaign, the continuum is near its brightest level, and it shows strong intensity variations. In contrast, the emission-line light curves do not rise to a peak until a third of the way through the campaign, starting off at rather low levels, and gradually rising in flux. Also, despite the significant peak in the first $\sim$50 days of the continuum light curve, the emission-line light curves are all relatively weaker. However, later in the campaign there are significant variations in line flux that do correlate with the continuum - the early emission-line light curves are not simply a smoothed, scaled, and shifted version of the continuum light curve. After the first third of the campaign, however, the emission lines track and correlate more closely with the continuum variations. Finally, during the last days of the campaign (2022 February 05 to 2022 February 24; HJD--2450000 9616--9634), the emission-line variations seem to decouple entirely from the continuum variations. This effect seems to be more prominent in the \ion{C}{4} emission-line light curve. 
This is possibly due to the appearance of a ``BLR holiday" \citep{Goad2016} at the end of the campaign. Even though the end of \hst\ UV observations limited investigation of this event, the ground-based spectroscopic campaign is still ongoing and will search for evidence of a BLR holiday at optical wavelengths. A thorough investigation of the light-curve behavior and individual reverberation windows will be discussed in an accompanying paper (Homayouni et al. 2022, in prep.) The reverberation lags between UV continuum bands will be described in a future paper in this series based on the combined HST and Swift data. In the next section we analyze these light-curve properties more quantitatively. Note that the analysis below is focused on the HJD--2450000 9177--9615 time range and excluding the last nine epochs that might be affected by a BLR holiday. 

\begin{figure*}[tt]
\centering
\includegraphics[width=0.98\textwidth]{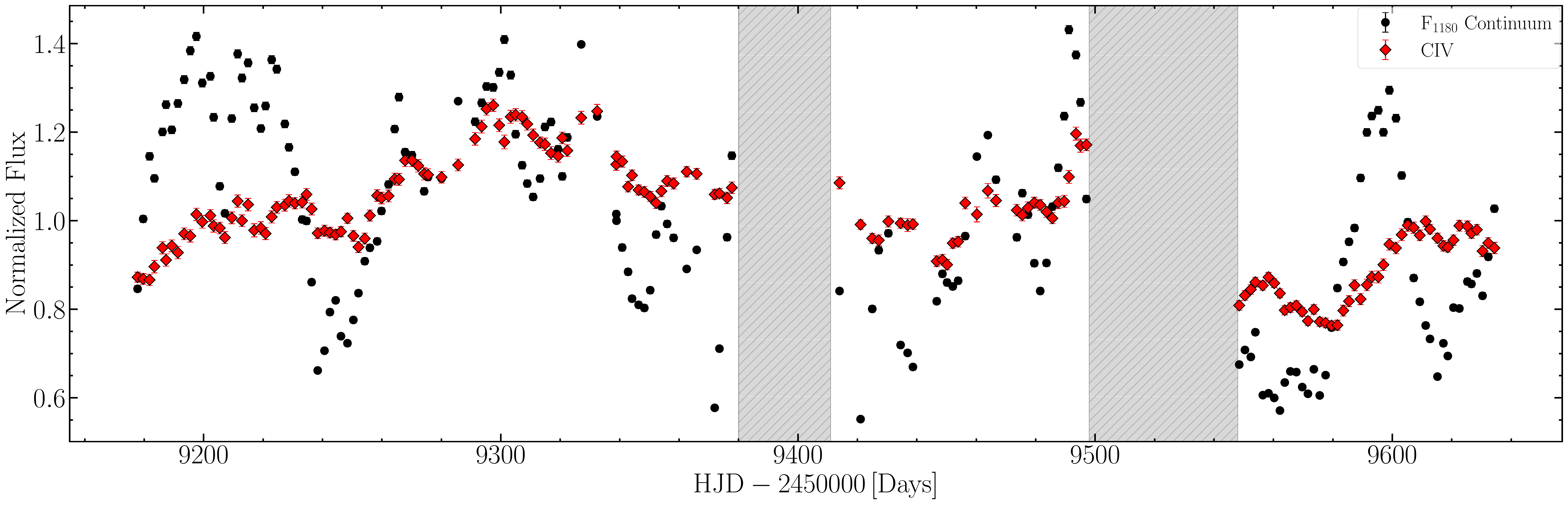}
 \caption{The 1180 \AA\ continuum light curve (black) superposed on the \ion{C}{4} light curve (red). Each light curve is normalized to its median. Treating the year-long duration of the campaign as a single light curve reveals a time-delay of $11.8_{-2.8}^{+3.0}$~days. However, the superposition of the continuum and line light curves show that a single time delay is an inadequate description of the light curve behavior. In particular, note the dramatic continuum variability near HJD 2459230 and HJD 2459590 days, where the strong features in the light curves indicate a lag $>$ 10 days, while away from these time periods, the light curves are more consistent with lags $<$ 10 days.}
\label{fig5_lc_overlap}
\end{figure*}

\startlongtable
\begin{deluxetable*}{cccccc}
\tabletypesize{\scriptsize}
\tablecaption{Integration Limits for Light Curves \label{tab:integration}}
\tablehead{
\colhead{Component} & \colhead{Integration Limit} & \colhead{Shortward Continuum} & \colhead{Longward Continuum} & \colhead{Total Line Flux Fraction ${}^{*}$} & \colhead{Velocity Range}\\
\colhead{} & \colhead{\AA} & \colhead{\AA} & \colhead{\AA} & \colhead{} & \colhead{$\mathrm{km \, s^{-1}}$}
} 
\startdata
$F_{\lambda}$(1180) & 1173.0--1185.0 & ... & ... & ... & ... \\
$F_{\lambda}$(1398) & 1394.2--1401.6 & ... & ... & ... & ... \\
$F_{\lambda}$(1502) & 1493.0--1511.0 & ... & ... & ... & ... \\
$F_{\lambda}$(1739) & 1736.0--1741.5 & ... & ... & ... & ... \\
\Lya\ & 1235.0--1248.0 & 1173.0--1185.0 & 1318.0--1330.0 & 21\% & -1500 to -4500\\ 
\ion{N}{5} & 1268.0--1290.0 & 1173.0--1185.0 & 1318.0--1330.0 & 64\% & -2250 to +2750\\ 
\ion{Si}{4} + \ion{O}{4} & 1424.0--1460.0 & 1394.2--1401.6 & 1493.0--1511.0 & 77\% & -4250 to +3000\\ 
\ion{C}{4} & 1590.0--1638.0 & 1493.0--1511.0 & 1736.0--1741.5 & 63\% & -1250 to +7500\\ 
\ion{He}{2} & 1680.0--1700.0 & 1493.0--1511.0 & 1736.0--1741.5 & 60\% & -2000 to +1500\\
\enddata
\tablecomments{
\footnotesize
$^{*}$ The total line flux fraction was obtained by integrating over the model of the mean spectrum for each emission line.
}
\end{deluxetable*}

\startlongtable
\begin{deluxetable*}{ccccccc}
\tablecaption{Light-Curve Statistics \label{tab:stats}}
\tablehead{
\colhead{Component} & \colhead{Mean and RMS flux} & \colhead{Mean Fractional Error} & \colhead{F$_{\rm Var}$$^{\star}$} & \colhead{Maximum Flux} & \colhead{Minimum Flux} & \colhead{Ratio$^{\dagger}$}\\
}
\startdata
$F_{\lambda}$(1180) & 76.73 $\pm$ 17.69 & 0.01 & 0.23 & 109.99 & 42.4 & 2.59 $\pm$ 0.03 \\ 
$F_{\lambda}$(1398) & 62.62 $\pm$ 12.74 & 0.01 & 0.2 & 88.74 & 38.88 & 2.28 $\pm$ 0.02 \\
$F_{\lambda}$(1502) & 57.22 $\pm$ 11.63 & 0.01 & 0.2 & 79.84 & 34.36 & 2.32 $\pm$ 0.02 \\
$F_{\lambda}$(1739) & 46.35 $\pm$ 9.35 & 0.03 & 0.2 & 68.25 & 28.01 & 2.44 $\pm$ 0.11 \\ 
\Lya\ & 11.83 $\pm$ 2.04 & 0.01 & 0.17 & 16.18 & 7.6 & 2.13 $\pm$ 0.03 \\ 
\ion{N}{5} & 16.53 $\pm$ 2.97 & 0.01 & 0.18 & 22.3 & 10.33 & 2.16 $\pm$ 0.03 \\ 
\ion{Si}{4} + \ion{O}{4]} & 6.11 $\pm$ 1.19 & 0.02 & 0.19 & 8.48 & 3.51 & 2.42 $\pm$ 0.07 \\  
\ion{C}{4} & 23.21 $\pm$ 2.7 & 0.01 & 0.12 & 29.06 & 17.59 & 1.65 $\pm$ 0.03 \\  
\ion{He}{2} & 2.77 $\pm$ 0.49 & 0.07 & 0.16 & 3.94 & 1.67 & 2.36 $\pm$ 0.25 \\ 
\enddata
\tablecomments{
\footnotesize
All light-curve statistics are reported in the observed frame. The continuum light curve at $F_{1180}$ is in units of $10^{-15}\mathrm{\,erg\, s^{-1}\, cm^{-2} \AA^{-1}}$ and the line fluxes are in units of $10^{-13}\mathrm{\,erg\, s^{-1}\, cm^{-2}}$.\\
$^{\star}$F$_{\rm Var}$ is defined as $({\sigma^2 - \delta^2})^{1/2}/{\left\langle F \right\rangle}$, where $\left\langle F \right\rangle$ is the mean of the observed flux, $\sigma$ is the RMS of the observed flux (second column), and $\delta$ is the mean statistical uncertainty (mean fractional error $\times$ $\left\langle F \right\rangle$).\\
$^{\dagger}$ Ratio is the maximum divided by the minimum of the observed flux.}
\end{deluxetable*}

\subsection{Unusual Light Curve Variability}
Figure~\ref{fig5_lc_overlap} shows the superposition of the 1180~$\AA$ continuum and \ion{C}{4} light curves. The continuum is initially strongly variable while the emission line light curves vary weakly and correlate poorly with the continuum; the continuum at the beginning of the campaign was  near its peak levels, but the emission lines were not. This weak response and lower amplitude of variations is reminiscent of the ``BLR Holiday" period of the STORM 1 campaign \citep{Goad2016, Dehghanian2019a}. However, after HJD 2459232 the emission line light curves became well-correlated with the continuum. Similar periods of weak correlation reverting to a strong response occurred throughout the remaining year of the campaign. These changes in the emission line response to continuum variations leads to complexities in the lag measurement. In fact, the lag analysis presented in \citet{Kara2021} shows lags that are shorter than the measurements reported in this work (see Section~\ref{sec:Analysis}). However, as we will discuss in upcoming work, the measured lag is a strong function of time interval analyzed. Further analysis of the light curve variability and implications for the time delay is beyond the scope of this work, and will be presented in a subsequent paper.

\section{Time-Series Analysis} \label{sec:timeseries}

We use several widely-used time-series analysis methods for measuring reverberation lags: \pyccf\ \citep{Sun2018b}, \texttt{ZDCF} \citep{Alexander2013}, and \jav\ \citep{Zu2011}. Below, we provide a brief overview of these lag measurement methods.

\subsection{ICCF}\label{sec:iccf}
The Interpolated Cross Correlation Function (ICCF) has been the most common method applied to previous RM studies \citep{Gaskell1986, Gaskell1987, Peterson2004}. ICCF calculates the Pearson coefficient $r$ between two light curves shifted by a range of time lags ($\tau$), using linear interpolation to match the shifted light curves in time, measuring $r$ over the range of allowed $\tau$. The centroid of the ICCF is computed using points surrounding the maximum correlation coefficient $r_{\mathrm{max}}$ ($r>0.8\, r_{\mathrm{max}}$), and the lag uncertainty is computed from Monte Carlo (MC) iterations of flux resampling and random subset sampling \citep{Peterson1998}. We implement ICCF using the publicly available \pyccf\ code \citep{Sun2018b} with an interpolation step of 1 day and a lag search range of $\pm$50 days. To estimate the uncertainties, we use 20000 Monte Carlo iterations of flux resampling and random subset sampling for \pyccf\ to generate a cross-correlation centroid distribution (CCCD), which gives the distribution of measured lags in all of the MC realizations. The upper and lower lag uncertainties are measured from the 16th and 84th percentiles of the primary peak. We report the lag centroid from the \pyccf\ method in Table \ref{tab:lags}.
We also tabulate the full-width at half-maximum (FWHM) for each emission line obtained from our fit to the mean spectrum. The FWHM is measured by taking the peak relative to the continuum-subtracted line profile and measure the Full Width at Half Maximum without excluding the negligible narrow-line contribution.

\startlongtable
\begin{deluxetable*}{ccccc}
\tablecaption{Continuum Light Curves \label{tab:lcs_con}}
\tablehead{
\colhead{HJD--2450000} & \colhead{$F_{\lambda}$ (1180 \,$\mathrm{\AA}$)} & \colhead{$F_{\lambda}$ (1398 \,$\mathrm{\AA}$)} & \colhead{$F_{\lambda}$ (1502 \,$\mathrm{\AA}$)} & \colhead{$F_{\lambda}$ (1739 \,$\mathrm{\AA}$)} \\
}
\startdata
9177.7825 & 65.01 $\pm$ 0.56 & 52.26 $\pm$ 0.34 & 48.07 $\pm$ 0.27 & 39.15 $\pm$ 1.20 \\ 
9179.6703 & 77.14 $\pm$ 0.61 & 60.50 $\pm$ 0.37 & 54.52 $\pm$ 0.28 & 44.66 $\pm$ 1.27 \\ 
9181.8164 & 88.00 $\pm$ 0.65 & 68.33 $\pm$ 0.39 & 60.57 $\pm$ 0.30 & 50.26 $\pm$ 1.33 \\ 
9183.4419 & 84.18 $\pm$ 0.64 & 67.42 $\pm$ 0.39 & 59.37 $\pm$ 0.29 & 50.57 $\pm$ 1.33 \\ 
9186.1714 & 92.24 $\pm$ 0.66 & 72.19 $\pm$ 0.40 & 63.84 $\pm$ 0.30 & 50.31 $\pm$ 1.33 \\ 
9187.3528 & 96.96 $\pm$ 0.68 & 75.47 $\pm$ 0.41 & 66.23 $\pm$ 0.31 & 55.85 $\pm$ 1.38 \\ 
9189.2933 & 92.61 $\pm$ 0.66 & 72.02 $\pm$ 0.40 & 65.39 $\pm$ 0.31 & 51.66 $\pm$ 1.34 \\ 
9191.3455 & 97.19 $\pm$ 0.68 & 75.12 $\pm$ 0.41 & 66.27 $\pm$ 0.31 & 55.56 $\pm$ 1.38 \\ 
9193.4633 & 101.32 $\pm$ 0.69 & 76.88 $\pm$ 0.41 & 69.08 $\pm$ 0.32 & 54.77 $\pm$ 1.38 \\ 
9195.5147 & 106.34 $\pm$ 0.71 & 81.24 $\pm$ 0.42 & 71.99 $\pm$ 0.32 & 58.40 $\pm$ 1.41 \\ 
\enddata
 \tablecomments{
\footnotesize{The continuum light curves are in units of $10^{-15} \mathrm{\,erg\, s^{-1}\, cm^{-2} \AA^{-1}}$. The full machine-readable table is available in the online journal article.}}
\end{deluxetable*}

\vspace{-0.2in}

\startlongtable
\begin{deluxetable*}{cccccc}
\tablecaption{Emission-Line Light Curves \label{tab:lcs_line}}
\tablehead{
\colhead{HJD--2450000} & \colhead{F{ (\Lya)}} & \colhead{F{ (\ion{N}{5}})} & \colhead{F{( \ion{Si}{4} + \ion{O}{4]})}} & \colhead{F{ (\ion{C}{4}})} & \colhead{F{ (\ion{He}{2}})}\\
}
\startdata
9177.7825 & 9.32 $\pm$ 0.10 & 14.39 $\pm$ 0.13 & 5.15 $\pm$ 0.10 & 20.11 $\pm$ 0.26 & 2.22 $\pm$ 0.17 \\ 
9179.6703 & 9.49 $\pm$ 0.11 & 14.41 $\pm$ 0.14 & 5.17 $\pm$ 0.11 & 20.02 $\pm$ 0.28 & 2.28 $\pm$ 0.18 \\ 
9181.8164 & 10.02 $\pm$ 0.11 & 14.41 $\pm$ 0.15 & 4.88 $\pm$ 0.12 & 19.97 $\pm$ 0.29 & 2.29 $\pm$ 0.20 \\ 
9183.4419 & 10.23 $\pm$ 0.11 & 15.10 $\pm$ 0.14 & 5.06 $\pm$ 0.12 & 20.66 $\pm$ 0.32 & 2.25 $\pm$ 0.21 \\ 
9186.1714 & 10.50 $\pm$ 0.11 & 14.91 $\pm$ 0.15 & 5.13 $\pm$ 0.12 & 21.64 $\pm$ 0.31 & 2.82 $\pm$ 0.20 \\ 
9187.3528 & 10.63 $\pm$ 0.11 & 15.10 $\pm$ 0.16 & 5.33 $\pm$ 0.12 & 21.00 $\pm$ 0.32 & 2.29 $\pm$ 0.21 \\ 
9189.2933 & 10.25 $\pm$ 0.12 & 15.05 $\pm$ 0.15 & 5.33 $\pm$ 0.12 & 21.74 $\pm$ 0.31 & 2.56 $\pm$ 0.21 \\ 
9191.3455 & 10.76 $\pm$ 0.12 & 15.41 $\pm$ 0.15 & 5.34 $\pm$ 0.12 & 21.39 $\pm$ 0.32 & 2.04 $\pm$ 0.20 \\ 
9193.4633 & 10.60 $\pm$ 0.12 & 15.82 $\pm$ 0.16 & 5.53 $\pm$ 0.12 & 22.37 $\pm$ 0.31 & 2.64 $\pm$ 0.20 \\ 
9195.5147 & 10.99 $\pm$ 0.12 & 15.93 $\pm$ 0.15 & 5.75 $\pm$ 0.13 & 22.26 $\pm$ 0.32 & 2.66 $\pm$ 0.21 \\ \enddata
 \tablecomments{
\footnotesize{The line fluxes are in units of $10^{-13} \mathrm{\,erg\, s^{-1}\, cm^{-2}}$. The full machine-readable table is available in the online journal article. The flux measurements reported here are only a fraction of the total line flux due to contamination by absorption and blending (see the reported fractions in Table~\ref{tab:integration}). }}
\end{deluxetable*}

\begin{figure}
\centering
\includegraphics[width=0.48\textwidth]{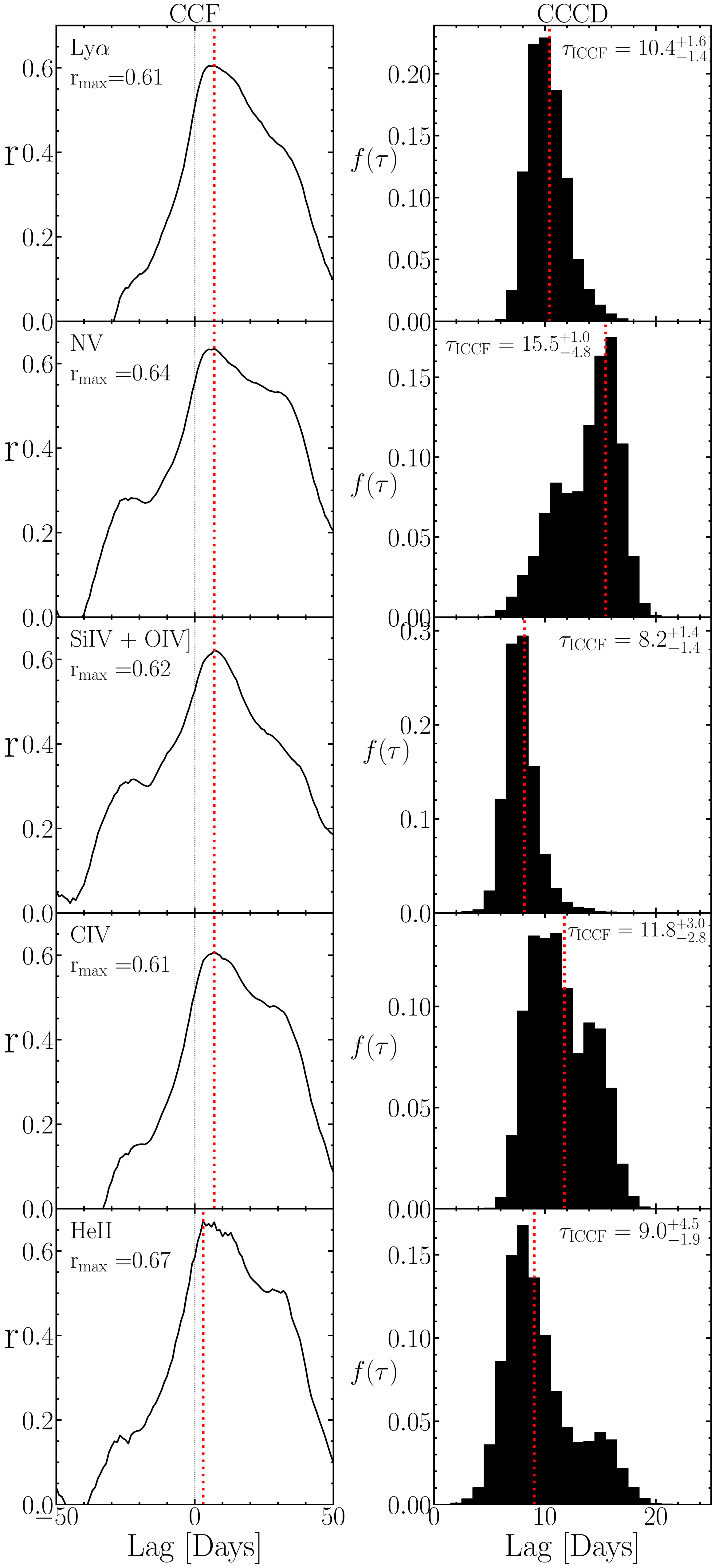}
\caption{Results of the lag analysis using \pyccf\ for the emission-line light curves. For each $F_{1180}$ and emission-line light-curve pair, we compute the cross-correlation coefficient with its maximum displayed by a red vertical line. The \pyccf\ cross-correlation centroid distributions (CCCD) are shown as black histograms, where $\mathbf{f(\tau)=N(\tau)/20000}$ is the total number of MC iterations normalized to unity. The maximum correlation coefficient and the lag measurement and uncertainties are displayed at the top-left corner of each panel.
}
\label{fig6:cccd}
\end{figure}

In this work, we only consider the continuum and line light curves for 2020 November 24 to 2022 February 05 (HJD 2459177 to 2459615), and exclude the last nine epochs from our lag analysis. We report the full campaign rest-frame time-delays and uncertainties in Table~\ref{tab:lags} for all UV emission lines as indicators of the characteristic sizes of the regions that produced the observed emission line flux variations.
Figure \ref{fig6:cccd} shows cross-correlation distribution functions from the \pyccf\ analysis along with the MC distribution of lag measurements for each emission line.
Note that although all the CCFs have strong peaks with $ r > 0.5$, they are very broad, likely indicating that the emission-line regions are spread over a large range in radial distance.
Furthermore, our lag analysis reveals that the lags of the CCF peaks are consistently shorter than the CCF centroid lags for all the UV emission lines (see Table~\ref{tab:lags}). This is caused by the asymmetry in the CCF, which may be an indication of an asymmetric transfer function or emission from an extended region \citep{Cackett2007, Cackett2022}.
We emphasize that the reported time delays in Table~\ref{tab:lags} are based on a single treatment of the light curve in the reverberation mapping scheme. Given the width and complexity of the CCFs, such a simple interpretation may not be an accurate description of the emission-line response to the varying continuum over the year-long campaign. We will discuss the implications of this behavior in an accompanying work (Homayouni et al. 2022 in prep.)

\subsection{Alternate Methods of Lag Measurement}
In order to ensure that the time-delay measurements are not due to our choice of lag measurement method, we additionally used the \texttt{ZDCF} (Z-transformed discrete cross-correlation function; \citealp{Alexander2013}) approach, in combination with a Gaussian process regression (GP), to model the stochastic AGN light curves with arbitrary sampling (see \citealt{Pancoast2015, Grier2017, Kovacevic2018}). For details on the approach, see \citet{Kovacevic2017}. We find that \texttt{ZDCF} yields similar time-delay measurements as \pyccf\ with time delays consistent within $1\sigma$.

We also investigated whether a long-term trends in the light curve could affect these time-delay measurements \citep{Welsh1999}. We used a running-median approach to remove the long-term trend and ``detrended" the light curves. Using the detrended light curves, we measured the time delays. However, we found that a single trend covering the entire duration of our campaign does not fully capture the variability of the light curves. We will address more complex ways of detrending in a parallel work (Homayouni et~al. 2022, in prep.)

\begin{table}
\bigskip
\caption{Emission Line Lags and Widths}\label{tab:lags}
\begin{center}
  \begin{tabular}{ccccc}
    \hline
    \hline
    Emission Line & Lag Centroid  & Lag Peak & $r_{\rm max}$ & FWHM\\
    {} & {Days} & {Days} & {} & ($\rm km\,s^{-1}$)\\
    \hline
    \Lya & $10.4_{-1.4}^{+1.6}$ & $6.8_{-1.9}^{+2.9}$ & 0.61 & 3610\\ 
    \ion{N}{5} & $15.5_{-4.8}^{+1.0}$ & $6.8_{-1.9}^{+1.9}$ & 0.64 & 3490\\ 
    \ion{Si}{4} + \ion{O}{4]} & $8.2_{-1.4}^{+1.4}$ & $6.8_{0.0}^{+1.9}$ & 0.62 & 5220\\ 
    \ion{C}{4} & $11.8_{-2.8}^{+3.0}$ & $ 6.8_{-1.9}^{+2.9}$ & 0.61 & 3750\\ 
    \ion{He}{2} & $9.0_{-1.9}^{+4.5}$ & $ 2.9_{0.0}^{+8.7}$ & 0.67 & 7830\\   
    \hline
 \end{tabular}
\end{center}
\tablecomments{
\footnotesize
The reported rest-frame time delays are based on the full campaign light curve (HJD$-$2450000=9177--9615) with a single lag measurement, and thus the complex response of the emission line responses to the continuum light curve might not be sufficiently captured by it. The last column, FWHM, gives the full-width at half-maximum of each emission line in the mean spectrum.
}
\end{table}
Finally, we also measure time delays using \jav\ \citep{Zu2011}. \jav\  uses a damped random walk (DRW) model to describe the stochastic variability of the AGN light curves. \jav\ relies on the underlying assumption of physical reverberation: the emission-line light curves are scaled, time-delayed, and smoothed versions of the continuum light curve.
\jav\ uses a Markov chain Monte Carlo approach using a maximum-likelihood method to fit a DRW model to the continuum and emission-line light curves, assuming that the local accretion-disk response is a top-hat function and the reverberating light curve model is the smoothed, scaled, and shifted version of the continuum light curve. 

We allow the DRW amplitude to be a free parameter but fix the DRW damping timescale to 2000 days. Our campaign duration ($\sim$450 days) is much smaller than the typical damping timescale of an AGN ($\sim$1500 days in the observed frame; see \citealp{Kelly2009, MacLeod2012}). Thus, the damping timescale's exact value does not matter, so long as it is longer than the campaign's duration (the light curves are effectively modeled as a red-noise random walk with minimal damping). We also tested damping timescales of 500, 1000, 5000, and 10,000 days. Similar to the \pyccf\ method (Section~\ref{sec:iccf}), we measured the emission-line lags relative to the continuum $F_{1180}$ variations. However, we found that a single DRW fit from \jav\ to the entire duration of our campaign does not fully capture the variability of the light curve because as noted earlier, the line light curves are not simply smoothed and shifted versions of the continuum. The resulting model light curves do not track the data well in several time intervals of our campaign.

\begin{figure*}
\centering
\includegraphics[width=0.9\textwidth]{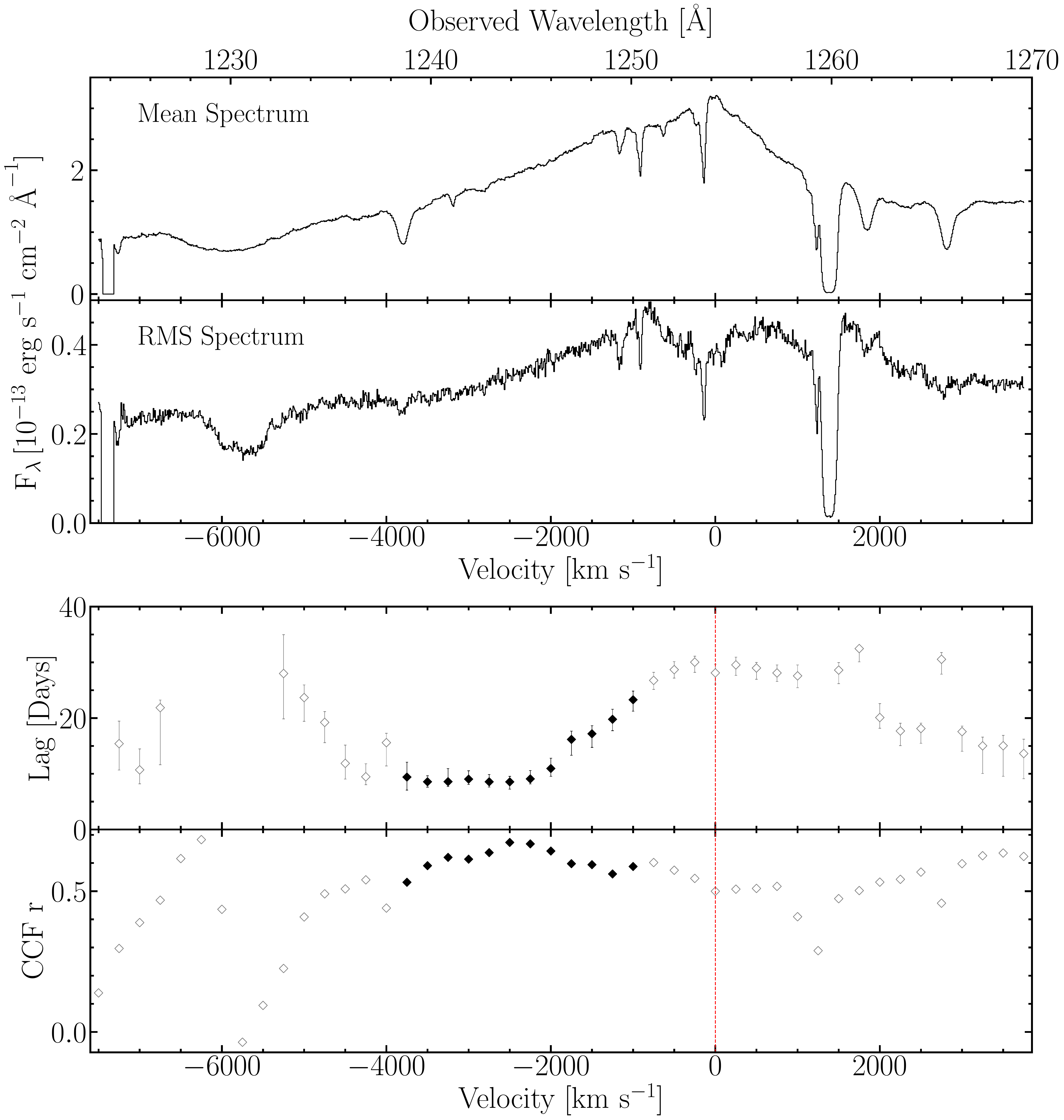}
 \caption{Rest-frame velocity-dependent time lags for the \Lya\ emission line. The top two panels show the \Lya\ mean and RMS line profile respectively. The bottom two panels show the distribution of velocity-binned lags for each 250 $\rm km\,\,s^{-1}$ bin and the associated maximum cross-correlation value $r$. The \Lya\ emission-line profile is highly contaminated by several broad absorption features both at the line center (due to \ion{N}{5}) and in the red wing (as indicated by open symbols). However, the blue wing shows a clear signature of virialized \Lya\ gas where high-velocity gas responds more rapidly, and low-velocity gas that is toward the line center responds later. Detailed absorption models are needed to extract the underlying \Lya\ emission-line profile and recover the velocity-dependent lag results, which will be the subject of future work in the series.}
\label{fig7_vel_res_lya}
\end{figure*}

\begin{figure*}
\centering
\includegraphics[width=0.9\textwidth]{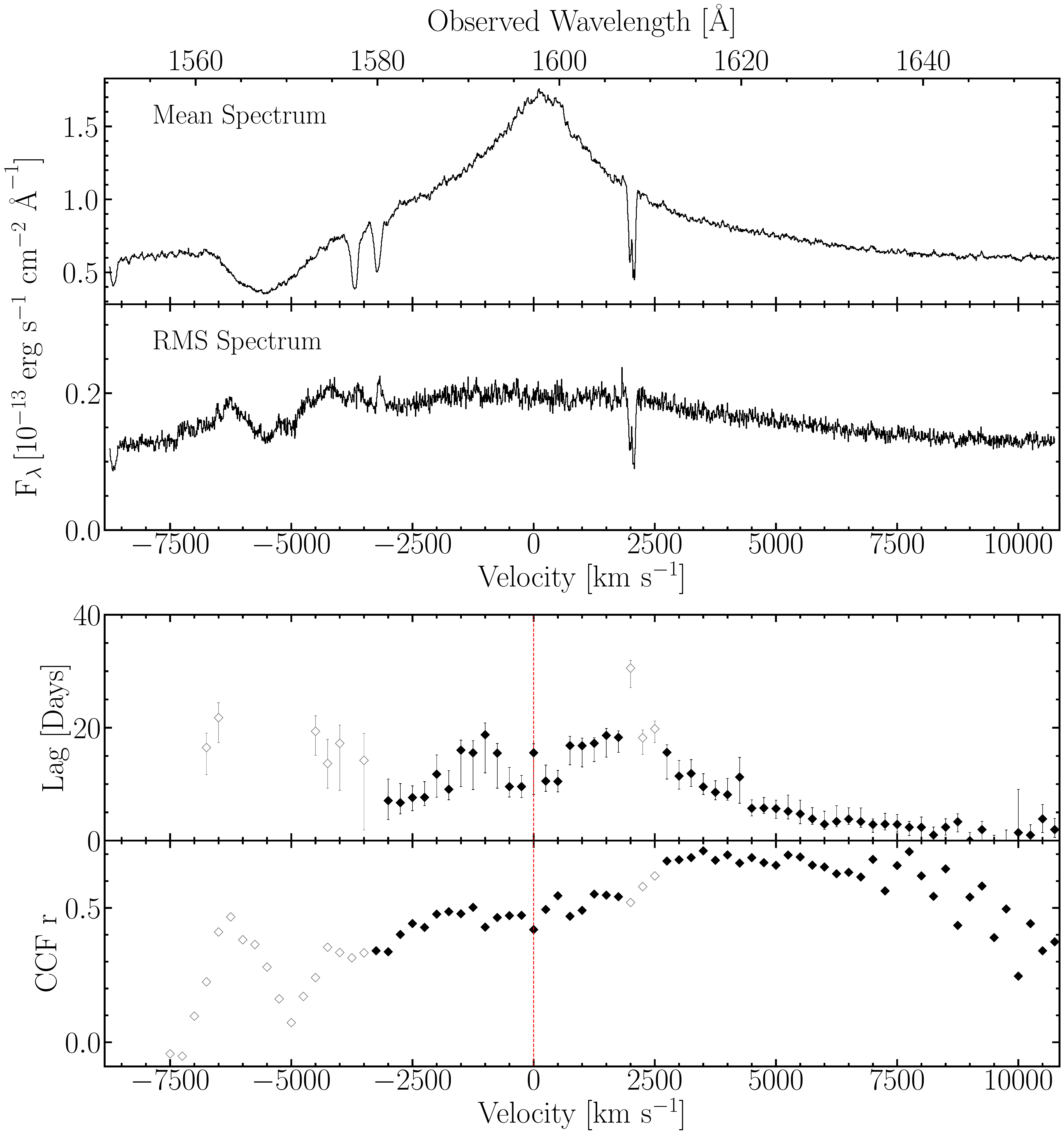}
 \caption{Rest-frame velocity-dependent time lags for the \ion{C}{4} emission line. The top two panels show the \ion{C}{4} mean and RMS line profile, respectively. The bottom two panels show the velocity-binned lags for each bin and the associated of the maximum cross-correlation coefficient $r$. The open symbols illustrate where the spectra has been contaminated by absorption. The lag velocity profile has an ``$M$" shape, with a local minimum near the line center, local peaks at $\sim\pm$ 1500 $\rm km\, s^{-1}$, and the shortest lags at higher velocity in the wings. }
\label{fig8_vel_res_civ}
\end{figure*}

\subsection{Velocity-Binned Results}\label{vel_res}
As discussed in Section~\ref{sec:design}, this program was designed to recover kinematic information about the BLR by resolving the emission-line response as a function of radial velocity. Similar to the AGN STORM~1 campaign on NGC\,5548 \citep{Horne2021}, the full analysis of velocity-delay maps will be presented in a subsequent paper in the AGN STORM~2 series. Here we carry out a preliminary analysis intended to show whether velocity-dependent information is present in the data. We isolate the \Lya\ and \ion{C}{4} profiles as described earlier in Section~\ref{sec:lightcurves} and integrate the fluxes in bins of width 250 $\mathrm{km\,s^{-1}}$, probing $-7500$ $\mathrm{km\,s^{-1}} < \Delta V <$ 4000 $\mathrm{km\,s^{-1}}$ for \Lya, and $-9000$ $\mathrm{km\,s^{-1}} < \Delta V <$ 11000 $\mathrm{km\,s^{-1}}$ for the \ion{C}{4} emission line. 

Figures~\ref{fig7_vel_res_lya} and \ref{fig8_vel_res_civ} show velocity-dependent lags for \Lya\ and \ion{C}{4} respectively. The \Lya\ emission line shows a strong velocity-dependent profile. However, blending with \ion{N}{5} on the red wing, and the presence of multiple broad absorption features, contaminate the \Lya\ emission-line flux and thus affect the results (see open symbols in Figure~\ref{fig7_vel_res_lya}). Many of the computed lags that are in absorption-contaminated bins are associated with small correlation coefficients and/or negative lags and are not included in the figure. We find small lags for the outer blue wing of the velocity profile, indicating that high-velocity gas responds more rapidly than the low-velocity gas near the line center. This general pattern is similar to what has been seen in other objects and is consistent with a virialized region. However, this virialized motion is mostly present in the central region of the \Lya\ profile, and the complex response of emission lines to continuum variations as characterized in Figures~\ref{fig4_lcs} and \ref{fig5_lc_overlap} could indicate that other gas motions may also be present in the BLR, but are less dominant near the center of the line profile. Full modeling of the spectra to remove the impact of absorption and the blending of the emission lines (see, e.g., STORM VIII \citealt{Kriss2019}), and different approaches to producing 2D velocity-resolved reverberation maps (see e.g., STORM IX \citealt{Horne2021}, and STORM XII \citealt{Williams2020}) will elucidate the gas dynamics in the BLR.

The \ion{C}{4} emission line shows strong velocity-dependent structure that spans a wide range of time-lags. We have similar problems due to absorption in the blue wing ($<1580$ \AA) of the \ion{C}{4} profile. The central portion of \ion{C}{4} shows a distinctive ``M"-shaped profile, characteristic of a thick rotating disk \citep{Welsh1991}.  The red wing, which is free of absorption, shows a virial profile with long lags near line center and shorter lags at higher velocities. To investigate the robustness of the velocity-binned lag results, we also show the maximum correlation coefficient profiles in the  \Lya\ and \ion{C}{4} emission-line profiles in the bottom panels of Figures \ref{fig7_vel_res_lya} and \ref{fig8_vel_res_civ}. Velocity bins showing poor correlation ($r<0.4$) are more frequent in windows that are affected by absorption. We need detailed modeling of the broad absorption features to recover the uncontaminated emission-line profiles and refine our velocity-binned analysis. This will be the subject of subsequent work in the AGN STORM~2 series of papers.

\section{Discussion}\label{sec:discussion}
Mrk~817 has been observed in previous ground-based optical campaigns \citep{Peterson1998, Denney2010} with multiple results for the \Hb\ mean lag and velocity-binned analysis. However, the current campaign provides the only velocity-binned results on Mrk~817 in the UV. Even though the target was primarily selected because of absence of absorption in its emission-line core and wings, the first \hst\ observation revealed the presence of strong broad and narrow absorption lines, which complicates measurement of the true emission-line fluxes. Nevertheless, we find significant variability in the continuum and emission-line light curves. We are able to measure BLR lags for prominent high-ionization lines that have not been observed before in \mrk.

Our analysis of the velocity-binned lags reveals that the \ion{C}{4} line shows a clear velocity-resolved response. Although broad absorption contaminates the far-blue wing of the \ion{C}{4} profile, it does not overlap with the central portion of the emission line, where we recover a strong velocity-dependent signal.
The \ion{C}{4} line has an ``M"-shaped velocity profile, with a local minimum ($\sim$10~days) near the line center and peak at longer lags ($\sim$~20 days) at roughly $\pm$ 1500 $\rm km\, s^{-1}$, which is indicative of a thick rotating disk \citep{Welsh1991}. Similar profiles have been found in NGC~5548 for \Lya\ and \ion{C}{4} \citep{Kriss2019}, and also for H$\beta$ \citep{DeRosa2018}. 

The \Lya\ profile is highly contaminated by absorption troughs or blended with \ion{N}{5}. However, in the narrow, contamination-free window of the \Lya\ blue wing, we find kinematic signatures that are consistent with an extended BLR in virial motion.

Although we can measure time lags for all the prominent UV emission lines, the observed light curves are not simply the  smoothed, scaled, and shifted versions of the continuum light curve. This complicated response of the BLR to the continuum variations is likely associated with the broad and narrow absorption features in the UV spectra, and to the heavy obscuration in the X-rays \citep{Kara2021}. This absorbing material can shield the BLR from the ionizing continuum, making the observed UV continuum an imperfect proxy for the true ionizing flux. 

The AGN STORM~2 data includes UV emission lines with a wide range of ionization potentials and probe a broad range of lag measurements (Table~\ref{tab:lags}). Typically, we expect shorter lags for emission lines with higher ionization potentials. Examining the centroids of the lag distributions in Table~\ref{tab:lags}, however, we see that this sequence is completely reversed. Lines with the highest ionization potentials (\ion{N}{5} and \ion{He}{2}) have the longest lags, and lower-ionization ions such as \ion{Si}{4} + \ion{O}{4]} have the shortest lags. This unexpected relationship is another reflection of the complexity of the BLR structure we have uncovered in the campaign. However, the interplay between ionization state and thermalization density could introduce deviations from this expected relation. For example, despite its relative small formation ionization potential, \ion{Si}{4}~1398~\AA\ has a surprisingly small expected emissivity-weighted radius, comparable to that of \ion{O}{4]}~1402~\AA\ (see \citealp{Korista2019} for more detail). Similarly, \citet{Hu2020} found complicated structure in PG 2130+099 between the ionization potential and the lag for \ion{He}{1}, \Hb, and \ion{Fe}{2}.

In conclusion, the BLR gas distribution is more complicated than can be captured using a single lag measurement (see Figure~\ref{fig6:cccd}). The CCFs for all the lines in Figure \ref{fig6:cccd} are noticeably asymmetric, and the peaks of the distributions, also shown in Figure~\ref{fig6:cccd}, generally show shorter lags than the centroids. Using this measure, the shortest lag is for the high-ionization line \ion{He}{2}, with all other lines having only a slightly longer peak lag.
Given these complexities, we are not able to analyze fully the BLR ionization stratification in this work, and we leave a more detailed examination of the complex behavior of the BLR response to a parallel paper (Homayouni et al. 2022, in prep.) 

Preliminary results from this analysis show the potential to reveal the types of structured maps that will provide additional constraints on future models of the BLR, and more clearly reveal distinct kinematic structures responsible for the velocity-resolved signatures we presented here.

\section{Summary}
We present the first results from 165 epochs of \hst\ COS observations as part of the multiwavelength AGN STORM~2 campaign on Mrk~817 to measure mean and velocity-dependent RM time lags for broad UV emission lines. Mrk~817 was observed with an average 2-day cadence from 2020 November 24 to 2022 February 24 (HJD$-$245000=9177--9634). This work reports the final custom calibration of the COS data with homogeneous reduction that have a local flux precision $<$1.5\%. Our major findings are as follows.

\begin{enumerate}
    \item The UV continuum at 1180~\AA\ and the emission-line light curves for \Lya, \ion{N}{5}, \ion{Si}{4} + \ion{O}{4]}, \ion{C}{4}, and \ion{He}{2} show significant variability.
    \item HST observations from the last 9 epochs point to a plausible ``broad-line region holiday." Ongoing ground-based observations may be able to reveal whether the BLR holiday also extends to the optical regime.
    \item We report the mean time-delay measurements for all UV emission lines (see Table~\ref{tab:lags}) with respect to the 1180~\AA\ continuum. However, we find that a single lag measurement is an inadequate description of the light curves. The emission-line responses are not simply a smoothed, scaled, and shifted version of the continuum variations. We will discuss the complex behavior of the BLR response in an accompanying paper (Homayouni et al. 2022, in prep.)
    \item Our qualitative analysis of the velocity-dependent lags only focuses on \Lya\ and \ion{C}{4} owing to significant contamination of the rest of the emission-line profiles. Velocity-dependent lags in the blue wing of \Lya\ are consistent with virial motion, with the shortest lags present in the high-velocity wings, and the longest lags in the lower-velocity center. The velocity-resolved  analysis for \ion{C}{4} instead reveals an ``$M$"-shaped profile hinting at BLR gas in a thick rotating disk.  
    \item There is no simple connection between emission-line ionization potentials and their BLR lags. The BLR gas and the emission-line response are more complex than can be reflected by a single lag measurement. 
\end{enumerate}

We plan to use our custom-calibrated \hst\ light curves for the analyses in upcoming work. Even though the \hst\ portion of the campaign has ended, monitoring of Mrk~817 is continuing with Swift, NICER, and ground-based observations.

\begin{acknowledgments}
This paper is the second in a planned series of papers by the AGN STORM~2 collaboration. Our project began with the successful Cycle 28 HST proposal 16196 \citep{Peterson2020}. Support for Hubble Space
Telescope program GO-16196 was provided by NASA through
a grant from the Space Telescope Science Institute, which is operated by the Association of Universities for Research in Astronomy, Inc., under NASA contract NAS5-26555. We are grateful to the dedication of the Institute staff who worked hard to review and implement this program. We particularly thank the Program Coordinator, W. Januszewski, who made sure
the intensive monitoring schedule and coordination with other facilities continued successfully.

Y.H. acknowledges support from NASA grant HST-GO-16196, and was also supported as an Eberly Research Fellow by the Eberly College of Science at the Pennsylvania State University. Research at UC Irvine has been supported by NSF grant AST-1907290. Research at Wayne State University was supported by NSF grant AST 1909199, and NASA grants 80NSSC21K1935 and 80NSSC22K0089. E.K. acknowledges support from NASA grants 80NSSC22K0570 and GO1-22116X. M.C.B. gratefully acknowledges support from the NSF through grant AST-2009230. T.T. and P.R.W. acknowledge support by NASA through grant HST-GO-16196, by NSF through grant NSF-AST 1907208, and by the Packard Foundation through a Packard Research Fellowship to T.T. G.J.F. and M.D. acknowledge support by NSF (1816537, 1910687), NASA (ATP 17-ATP17-0141, 19-ATP19-0188), and STScI (HST-AR- 15018 and HST-GO-16196.003-A). P.B.H. is supported by NSERC grant 2017-05983. M.V. gratefully acknowledges financial support from the Independent Research Fund Denmark via grant number DFF 8021-00130. D.H.G.B. acknowledges CONACYT support \#319800 and of the researchers program for Mexico. D.C. acknowledges support by the ISF (2398/19) and the D.F.G. (CH71-34-3). A.V.F. was supported by the U.C. Berkeley Miller Institute of Basic Research in Science (where he was a Miller Senior Fellow), the Christopher R. Redlich Fund, and numerous individual donors.The UCSC team is supported in part by the Gordon and Betty Moore Foundation, the Heising-Simons Foundation, and by a fellowship from the David and Lucile Packard Foundation to R.J.F.
\end{acknowledgments}

\bibliography{main.bib}

\end{CJK*}  
\end{document}